\documentclass[usenatbib,useAMS,letterpaper]{mn2e}
\usepackage{float}
\usepackage{graphicx}
\usepackage{epstopdf}
\usepackage{ctable}
\usepackage{url}
\usepackage{times}
\usepackage{amsmath}
\usepackage{amssymb}
\usepackage{xspace}
\usepackage{mathrsfs} 
\usepackage{tabularx}
\usepackage{fixltx2e}
\usepackage{color}
\usepackage{placeins}

\newcommand{\acknowledgments}{\begin{small}\section*{Acknowledgements}\end{small}}

\newcommand\sref[1]{\hyperref[#1]{\S~\ref*{#1}}}
\newcommand\fref[1]{\hyperref[#1]{Fig.~\ref*{#1}}}
\newcommand\Eqref[1]{equation~(\hyperref[#1]{\ref*{#1}})}
\newcommand\tref[1]{\hyperref[#1]{Table~\ref*{#1}}}
\newcommand\aref[1]{\hyperref[#1]{Appendix~\ref*{#1}}}

\title[Feedback and microphysics in galaxy formation]{Feedback first: the surprisingly weak effects of magnetic fields, viscosity, conduction, and metal diffusion on sub-L$^*$ galaxy formation}
\author[K. Su et al.]{
\parbox[t]{\textwidth}{
Kung-Yi Su$^{1}$\thanks{E-mail: ksu@caltech.edu}, Philip F. Hopkins$^{1}$, Christopher C. Hayward$^{2,1,3}$, Claude-Andr\'e Faucher-Gigu\`ere$^4$, Du\v san Kere\v s$^5$,  Xiangcheng Ma$^{1}$, Victor H. Robles$^{6}$
}
\vspace*{6pt} \\
$^1$TAPIR 350-17, California Institute of Technology, 1200 E. California Boulevard, Pasadena, CA 91125, USA\\
$^2$Center for Computational Astrophysics, Flatiron Institute, 162 Fifth Avenue, New York, NY 10010, USA\\
$^3$Harvard-Smithsonian Center for Astrophysics, 60 Garden Street, Cambridge, MA 02138, USA\\
$^4$Department of Physics and Astronomy and CIERA, Northwestern University, 2145 Sheridan Road, Evanston, IL 60208, USA\\
$^5$Department of Physics, Center for Astrophysics and Space Sciences, University of California at San Diego, 9500 Gilman Drive, La Jolla, CA 92093, USA\\
$^6$Center for Cosmology, Department of Physics and Astronomy, University of California, Irvine, CA 92697, USA}
\usepackage[breaklinks,colorlinks,citecolor=blue]{hyperref}
\usepackage[all]{hypcap}

\begin{document}
\long\def\/*#1*/{}
\date{Accepted by  MNRAS  (2017  June 13)}

\pagerange{\pageref{firstpage}--\pageref{lastpage}} \pubyear{2017}

\maketitle

\label{firstpage}

\begin{abstract}
Using high-resolution simulations with explicit treatment of stellar feedback physics based on the FIRE (Feedback in Realistic Environments) project, we study how galaxy formation and the interstellar medium (ISM) are affected by magnetic fields, anisotropic Spitzer-Braginskii conduction and viscosity, and sub-grid metal diffusion from unresolved turbulence.  We consider controlled simulations of isolated (non-cosmological) galaxies but also a limited set of cosmological
 ``zoom-in'' simulations. 
Although simulations have shown significant effects from these physics with weak or absent stellar feedback, the effects are much weaker than those of stellar feedback when the latter is modeled explicitly. The additional physics have no systematic effect on galactic star formation rates (SFRs) . In contrast, removing stellar feedback leads to SFRs being over-predicted by factors of $\sim 10 -100$. Without feedback, neither galactic winds nor volume filling hot-phase gas exist, and discs tend to runaway collapse to ultra-thin scale-heights with unphysically dense clumps congregating at the galactic center. With stellar feedback, a multi-phase, turbulent medium with galactic fountains and winds is established. At currently achievable resolutions and for the investigated halo mass range $10^{10}-10^{13} M_{\odot}$, the additional physics investigated here (MHD, conduction, viscosity, metal diffusion) have only weak ($\sim10\%$-level) effects on regulating SFR and altering the balance of phases, outflows, or the energy in ISM turbulence, consistent with simple equipartition arguments. We conclude that galactic star formation and the ISM are  primarily governed by a combination of turbulence, gravitational instabilities, and feedback. We add the caveat that AGN feedback is not included in the present work.
\end{abstract}

\begin{keywords}
methods: numerical --- MHD --- conduction --- turbulence --- ISM: structure ---  ISM: jets and outflows
\end{keywords}

\vspace{-0.5cm}
\section{Introduction} \label{S:intro}
Feedback from stars is essential to galaxy evolution. In isolated galaxy simulations without strong stellar feedback, giant molecular clouds (GMCs) experience  runaway collapse, resulting in star formation rates (SFRs) orders-of-magnitude higher than observed \citep{2010MNRAS.409.1088B,2011MNRAS.413.2935D,2011IAUS..270..235H,2011ApJ...740...74K,2011ApJ...730...11T,2011MNRAS.417..950H}. 
  This is in direct contradiction with the observed Kennicutt-Schmidt (KS)  relation, which shows that the gas consumption time of a galaxy is roughly $\sim 50 -100$  dynamical times \citep{1998ApJ...498..541K,1974ApJ...192L.149Z,1997ApJ...476..166W,1999ARA&A..37..311E,2009ApJS..181..321E}.
   Cosmological simulations without strong feedback face a similar challenge. The efficiency of cooling causes runaway collapse of gas to high densities within a dynamical time, ultimately forming far too many stars compared to observations (\citealt{1996ApJS..105...19K,1999MNRAS.310.1087S,2000MNRAS.319..168C,2003MNRAS.339..289S,2009MNRAS.396.2332K}, and references therein). 

\vspace{0.1cm}
Recent years have seen great progress in modeling feedback on galaxy scales \citep{2000ApJ...545..728T,2007MNRAS.374.1479G,2009ApJ...695..292C,2012MNRAS.423.2374U,2011MNRAS.417..950H, 2012MNRAS.421.3488H,2012MNRAS.421.3522H,2015arXiv150900853A,2015arXiv151005644H}. In \cite{2011MNRAS.417..950H, 2012MNRAS.421.3488H}, a detailed feedback model including radiation pressure, stellar winds, supernovae and photo-heating was developed and applied to simulations of isolated galaxies. 
They showed that stellar feedback is sufficient to maintain a self-regulated multi-phase interstellar medium (ISM),
with global structure in good agreement with the observations.  GMCs survive several dynamical times and only turn a few per cent of their mass into stars, and the galaxy-averaged SFR agrees well with the observed Kennicutt-Schmidt (KS) law. 
These models were extended with numerical improvements and additional cooling physics, and then applied to cosmological ``zoom in'' simulations in the FIRE (Feedback In Realistic Environments) project\footnote{Project web site: http://fire.northwestern.edu.}. 
A series of papers, using the identical code and simulation set have demonstrated that these feedback physics successfully reproduce a wide range of observations, including star formation histories of galaxies \citep{2014MNRAS.445..581H}, time variability of star formation \citep[][]{2015arXiv151003869S}, galactic winds \citep[][]{2015MNRAS.454.2691M},  HI content of galaxy halos (\citealt{2015MNRAS.449..987F,2016arXiv160107188F}; Hafen et al., in prep.), and galaxy metallicities \citep[]{2015arXiv150402097M}. 
Other groups (e.g. \citealt{2013MNRAS.428..129S}, who implemented energy injection from SNe and an approximate treatment of UV radiation pressure, and \citealt[e.g.,][]{ 2015arXiv150900853A}, who included momentum injection from SNe, radiation pressure and stellar winds) have also found that stellar feedback can regulate galaxy SFRs and lead to realistic disc morphologies. 

However, several potentially important physical processes have not been included in most previous galaxy formation simulations.  Magnetic fields have long been suspected to play a role in galaxy evolution because the magnetic pressure reaches equipartition with the thermal and turbulent pressures \citep{1996ARA&A..34..155B,2009ASTRA...5...43B}. Isolated galaxy simulations with magnetic fields -- but using more simplified models for stellar feedback -- have been studied in various contexts and suggest that magnetic fields can provide extra support in dense clouds, thus slowing down star formation  \citep{2009ApJ...696...96W,2012MNRAS.422.2152B,2013MNRAS.432..176P}.  Turbulent box simulations \citep{2005ApJ...629..849P,2007ApJ...663..183P} also suggest that  MRI-driven (magnetorotational instability) turbulence can suppress star formation at large radii in spiral galaxies. In particular, \cite{2015ApJ...815...67K} explicitly demonstrate such suppression from magnetic fields in a simulation of a turbulent box that includes momentum feedback from SNe.  Magnetic fields can also be important because of their effects on fluid mixing instabilities, including the Rayleigh-Taylor (RT) and Kelvin-Helmholtz (KH) instabilities \citep{1995ApJ...453..332J, 2015MNRAS.449....2M,2016arXiv160805416A}. These instabilities can potentially affect galaxy evolution through processes  including the evolution of supernovae (SN) remnants \citep{1996ApJ...465..800J,1996ApJ...472..245J,1999ApJ...511..774J,2000ApJ...534..915T, 2015ApJ...815...67K}.

Another potentially important effect is viscosity, which has been more extensively studied in simulations of galaxy clusters. It has been suggested that viscosity can affect the turbulent motion of the intracluster medium (ICM) or circum-galactic medium (CGM) and affect the KH stability of various structures in the ICM \citep[][]{2007PhR...443....1M}. 
It has been shown in particular that viscosity may be important for the dynamics of bubbles in the ICM inflated by active galactic nucleus (AGN) feedback or bursts of SNe activity \citep{2005MNRAS.357..242R, 2006MNRAS.371.1025S}. 
 
Thermal conduction, which in the presence of magnetic fields is highly anisotropic, affects the stability of plasmas at both galactic and cluster scales \citep{2009ApJ...699..348S,2010ApJ...720..652S, 2012MNRAS.422..704P,2016arXiv160805416A, 2012ApJ...747...86C} and the survival and mixing of multi-phase fluids. Combined with the effect of magnetic fields, conduction may be critical to determine the survival of cool clouds in galactic winds. 

Turbulent metal diffusion due to small-scale (un-resolvable) eddies may also have important effects. It has been suggested, for example, that unresolved turbulence in galaxy simulations may be important to effectively ``diffuse'' metals in the ISM and intergalactic medium \citep[IGM; e.g.,][]{2010MNRAS.407.1581S}, leading non-linearly to different cooling physics at halo centers and within the dense ISM. 

While most previous studies considered these physics in isolation, their effects and relative importance may be quite different in a realistic multi-phase ISM shaped by strong stellar feedback processes. 
Another challenge is that conduction and viscosity in magnetized plasmas are inherently anisotropic. Properly treating this anisotropy requires MHD simulations and is numerically non-trivial; consequently, most previous studies on galactic scales have considered only isotropic conduction and viscosity. However, studies which correctly treat the anisotropy have shown that this anisotropy can produce orders-of-magnitude differences and, in some cases, qualitatively different behavior \citep{2009ApJ...704.1309D,2015ApJ...798...90Z,2009ApJ...699..348S,2010ApJ...720..652S,2012ApJ...747...86C}

In this paper, we study the effects of these different microphysics in the presence of explicit models for stellar feedback. 
While the simulations analyzed here implement the same stellar feedback physics from the FIRE cosmological simulations, we focus primarily on non-cosmological simulations of isolated galaxies, because this allows us to achieve higher spatial and mass resolution, and to have well-controlled experiments with identical galaxy initial conditions. In cosmological runs, on the other hand, the inherently chaotic nature of the problem makes detailed one-to-one comparison of simulations with varied physics more complicated; we do, however, include a limited subset of these experiments. We also make use of a new, more accurate hydrodynamic solver, needed to properly treat MHD and anisotropic diffusion.

Overall, we find that at the resolutions currently achievable in isolated galaxy and cosmological simulations, MHD, anisotropic conduction and viscosity, and sub-grid turbulent metal diffusion play a relatively minor role in the regulation of star formation and of the phases and energetics of the ISM \emph{when the dominant effects of stellar feedback are simultaneously modeled}. 
We caution, however, that despite this result, some of these effects likely have some important and observationally interesting consequences on finer scales, such as for the survival of cool clouds in galactic winds \citep[e.g.,][]{2015MNRAS.449....2M, 2016arXiv160805416A,2016ApJ...822...31B}, and stellar abundance distribution patterns within star clusters or small galaxies. 
It is also possible that some important effects would only reveal themselves in simulations of much higher resolution than currently possible for galaxy simulations. 
Furthermore, the interaction of physical processes not included in our simulations with, e.g., magnetic fields is likely to prove important. 
This is the case in particular for the transport of cosmic rays, which a number of recent studies indicate may be an important form of feedback for galaxy evolution \citep[e.g.,][]{2012MNRAS.423.2374U, 2013ApJ...777L..16B, 2014ApJ...797L..18S, 2016arXiv160204856R,2016arXiv160407399P,2016MNRAS.462.2603P}.

The remainder of this paper is organized as follows: in \sref{S:methods}, we describe the initial conditions and the baryonic physics model of our default model. In \sref{s:add_ph}, we summarize the additional physics studied in this paper. In \sref{S:results}, we analyze the effects on the star formation histories, morphologies, phase structures, magnetic and turbulent energies, and outflows of our simulated galaxies. We discuss the reason why the fluid microphysics have minor effects in \sref{s:discussion} and conclude in \sref{S:conclusions}.

\vspace{-0.5cm}
\section{Methodology} \label{S:methods}

Our simulations use {\small GIZMO} \citep{2015MNRAS.450...53H} \footnote{A public version of this code is available at \href{http://www.tapir.caltech.edu/~phopkins/Site/GIZMO.html}{\textit{http://www.tapir.caltech.edu/$\sim$phopkins/Site/GIZMO.html}}.}, in its Meshless Finite Mass (MFM) mode. This is a mesh-free, Lagrangian finite-volume Godunov code designed to capture advantages of both grid-based and smoothed-particle hydrodynamics (SPH) methods built on the gravity solver and domain decomposition algorithms of {\small GADGET-3} \citep{2005MNRAS.364.1105S}. The numerical details of the hydrodynamic and MHD versions of the method are presented in \citet{2015MNRAS.450...53H}, \citet{2016MNRAS.455...51H}, and \citet{2015arXiv150907877H}. \cite{2016arXiv160207703H} present tests of the anisotropic diffusion operators used in our code. Extensive comparisons of dozens of test problems demonstrate good code behavior and convergence, in good agreement with state-of-the-art moving mesh codes (e.g.\ {\small AREPO}, \citealt{2010MNRAS.401..791S}) and grid codes (e.g.\ {\small ATHENA}, \citealt{2008ApJS..178..137S}), including on historically difficult problems such as those featuring the magneto-rotational instability (MRI), magnetic jet launching, and the KH and RT fluid-mixing instabilities. Convergence tests for our isolated galaxy simulations can be found in \aref{A:resolution}. 

Note that, for the sake of consistency,  previously published FIRE simulations (see references in \S~1) were run with the identical source code, using {\small GIZMO}'s ``P-SPH'' hydrodynamic solver. P-SPH is an SPH method with improvements designed to address some of the known shortcomings of SPH in treating e.g.\ fluid mixing instabilities \citep[see][]{2013MNRAS.428.2840H}. This was done to facilitate comparison by matching exactly the code used for the first FIRE paper, \citet{2014MNRAS.445..581H}, written before the MFM methods were developed. Unfortunately, as shown in \citet{2016MNRAS.455...51H} and \citet{2016arXiv160207703H}, P-SPH (while reasonably well-behaved on pure hydrodynamics problems) exhibits serious inaccuracies and may not converge on MHD and anisotropic diffusion problems. As a consequence, P-SPH {\em cannot} be used for our study here. We are therefore careful to distinguish our isolated galaxy simulations here from the primary ``FIRE project'' simulations, although they use the same (operator-split) code modules to treat stellar feedback. In fact, the updated code here - the ``FIRE - 2'' code, will be the subject of an extensive methods paper in preparation (Hopkins et al., in preparation) and was first used in \cite{2016ApJ...827L..23W} for studying the satellites around a Milky Way-mass galaxy. A detailed study of the effects of the hydrodynamic method and other numerical details on the conclusions from the previous FIRE simulations will be the subject of the methods paper.


\vspace{0cm}
\subsection{Initial conditions (ICs)}
\label{s:ic}
In this paper, five isolated (non-cosmological) galaxy models are studied to consider a range of characteristic galaxy types. Two cosmological zoom-in ICs are also included as a check that our conclusions are applicable in a fully cosmological environment. More details regarding the isolated disc galaxies and the cosmological simulations can be found in \cite{2011MNRAS.417..950H,2012MNRAS.421.3488H} and \cite{2014MNRAS.445..581H}, respectively, and are summarized in \tref{tab:ic} and below.  For all runs, a flat $\Lambda$CDM cosmology with $h=0.702$, $\Omega_M=1-\Omega_\Lambda=0.27$, and $\Omega_b=0.046$ is adopted.
 
Note that we have tested simulations with most of our ICs re-run at different resolution, with initial gas particle mass differing by a factor $\sim100$. Some absolute properties do vary according to the resolution. For example, finer ISM substructure is observed and some higher density regions are resolved as the resolution increases. Nonetheless, the main conclusions of this paper (the {\em relative} differences in runs with different microphysics) remain robust at all resolutions investigated. A detailed convergence study is presented in \aref{A:resolution}.

The ICs studied here include the following:

\begin{table*}
\begin{center}
 \caption{Galaxy models}
 \label{tab:ic}
 \begin{tabular*}{\textwidth}{@{\extracolsep{\fill}}cccccccccccccccc}
 \hline
\hline
Model          &$\epsilon_g$        &$m_g$                &$M_{\rm halo}$      &$c$    &$V_{\rm Max}$    &$M_{\rm bar}$       &$M_b$                &Bulge  &$a$      &$M_d$              & $r_d$      &$M_g$                &$r_g$     \\
                      &(pc)                        &(M$_\odot$)  &(M$_\odot$)   &           &(km/s)            &(M$_\odot$)   &(M$_\odot$) & profile &(kpc)   &(M$_\odot$)  &(kpc)       &(M$_\odot$)    &(kpc)    \\
\hline
HiZ                   &1.4                           &2.5e4                &2.1e12               &3.5      &280                 &1.53e11             &1.0e10                 &Exp                  &1.7        &4.3e10              &2.3            &1.0e11                   &4.6           \\
Sbc                 &1.4                           &2.6e3                  &2.1e11                &11      &120                    &1.5e10               &1.4e9                &Exp                  &0.5        &5.7e9                &1.9            &7.9e9                  &3.7           \\
MW                 &3.6                           &3.5e3                  &2.1e12                &12      &250                  &1.02e10             &2.1e10             &Hq       &1.4        &6.8e10              &4.3            &1.3e10               &8.6             \\
SMC               &0.7                           &3.6e2                     &2.9e10               &15       &67                    &1.3e9                  &1.4e7               &Hq       &2.1        &1.9e8                &1.0               &1.1e9                  &3.0             \\
Ell                    &4.2                           &7.1e4                 &1.4e13               & 6        &240                  &1.02e12             &1.4e11              &Hq       & 3.9       &1.4e10              &4.0            &8.6e11               &4.0                 \\
CosmoMW        &7                           &5.7e4                &1.2e12              & 8    & 290                  &1.3e11               &-                          &-                        &-            &1.2e11              &1.2            &7.1e9                  &2.5          \\
CosmoDwarf    & 3              & 2.6e2   & 7.9e9 &9.7  &20  &5.2e6  &-&-&-&1.7e6&-&3.5e6&-\\
\hline 
\hline
\end{tabular*}
\end{center}
\begin{flushleft}
Parameters of the galaxy models studied here (\sref{s:ic}):\\
(1) Model name. HiZ: high-redshift, massive starburst. Sbc: local gas-rich dwarf starburst. MW: Milky-Way analogue. SMC: SMC-mass dwarf. Ell: massive elliptical with an extended gaseous halo. HiZ, Sbc, MW, SMC, and Ell are non-cosmological (isolated galaxy) simulations. CosmoMW: cosmological simulation of a MW-mass disc galaxy. CosmoDwarf: cosmological simulation of a dwarf galaxy.
(2) $\epsilon_g$: Gravitational force softening for gas (the softening for gas in all simulations is adaptive; here, we quote the minimum Plummer equivalent softening for a kernel containing 32 particles).
(3) $m_g$: Gas particle mass.
(4) $M_{\rm halo}$: Halo mass. $M_{vir}$ for CosmoMW and CosmoDwarf.
(5) $c$: NFW-equivalent halo concentration.
(6) $V_{\rm max}$: Halo maximum circular velocity.
(7) $M_{\rm bar}$: Total baryonic mass. It is the sum of gas, disc, bulge and stellar mass for isolated galaxy runs, and the sum of gas and stellar mass in the cosmological runs within 0.1 virial radius.
(8) $M_b$: Bulge mass.
(9) Bulge profile: Hq: \cite{1990ApJ...356..359H}, or Exp: Exponential.
(10) $a$: Bulge scale-length.
(11) $M_d$ : Stellar disc mass. For CosmoMW and CosmoDwarf runs, this is the total stellar mass within 0.1 virial radius.
(12) $r_d$ : Stellar disc scale length.
(13) $M_g$: Gas disc mass. For the Ell  runs, this includes gas in the extended halo. For CosmoMW and CosmoDwarf runs, this is the total gas mass within 0.1 virial radius.
(14) $r_g$: Gas disc scale length.

The properties quoted for CosmoMW and CosmoDwarf are the $z=0$ values measured from the ``FB'' run. 
The CosmoDwarf does not have a well-defined disc even at $z=0$, but is a dwarf irregular galaxy.

\end{flushleft}
\end{table*}

\vspace{0cm}
\subsubsection{HiZ}
 HiZ is a high-redshift massive starburst disc galaxy designed to match the properties of non-merging, rapidly star-forming sub-millimeter galaxies \citep{2006ApJ...646..107E,2008ApJ...687...59G,2010Natur.463..781T}, with halo mass $M_{\rm halo}=2.1\times10^{12}$ M$_\odot$ in a \cite{1990ApJ...356..359H} profile with an NFW \citep{1996ApJ...462..563N}-equivalent concentration of $c=3.5$. The baryonic component has a total mass of $M_{\rm bar}=1.53\times10^{11}$ M$_\odot$ and consists of an exponential  bulge ($\rho(r)\propto \exp(- r/a)/r$) ($M_b=  10^{10}$ M$_\odot$) with  scale length $a=1.7$ kpc and exponential stellar ($M_d=4.3\times10^{10}$ M$_\odot$) and  gas ($M_g=1\times10^{11}$ M$_\odot$) discs with scale lengths $r_d=2.3$ kpc and $r_g=4.6$ kpc respectively. The gas disc initially has Toomre $Q=1$ uniformly. Note that the virial radius is scaled for a halo at redshift $z=2 $ instead of $z=0$. This model uses $1.65\times 10^{7}$ particles,  $4\times10^6$  of which are gas particles. The initial metallicity is set to $0.5 Z_{\odot }$\footnote{The solar metallicity, $Z_{\odot }$, of each species (Total, He, C, N, O, Ne, Mg, Si, S, Ca, Fe) is tabulated from \cite{2009ARA&A..47..481A}}.

\vspace{-0.5cm}
\subsubsection{Sbc}
Sbc is a $z=0$ dwarf starburst intended to be representative of local luminous infrared galaxies (LIRGs).  The IC is composed of a dark matter halo with $(M_{\rm halo},c)=(2.1\times 10^{11}$M$_\odot, 11)$ and  a baryonic component with masses $(M_{\rm bar}, M_b, M_d, M_g)=(15, 1.4, 5.7, 7.9)\times 10^9$M$_\odot$ and scale lengths $(r_d, r_g, a)=(1.9, 3.7, 0.5)$ kpc.  The bulge has an exponential profile. This model includes $1.7\times 10^7$ particles, $3\times 10^6$ of which are gas particles. The initial metallicity is set to $0.3 Z_{\odot }$.

\vspace{-0.5cm}
\subsubsection{MW}
MW is a Milky Way-like galaxy composed of a dark matter halo with $(M_{\rm halo},c)=(2.1\times 10^{12}$ M$_\odot,12)$ and baryonic components with $(M_{\rm bar},M_b,M_d,M_g)=(1.02,2.1,6.8,1.3)\times 10^{10}$M$_\odot$ respectively. The  scale lengths are $(r_d,r_g, a)=(4.3,8.6,1.4)$kpc. The bulge follows a \citet{1990ApJ...356..359H} profile. The model includes $1.03\times10^7$ particles, $3.6\times10^6$ of which are gas particles.  The initial metallicity is set to $Z_{\odot }$.

\vspace{-0.5cm}
\subsubsection{SMC}
SMC is an isolated (field)  Small Magellanic Cloud-mass dwarf galaxy composed of a halo with $(M_{\rm halo}, c)=(2.9 \times 10 ^{10}$ M$_\odot, 15)$ and baryonic components with masses $(M_{\rm bar}, M_b, M_d, M_g)=(13, 0.14, 1.9, 11)\times 10^8$ M$_\odot$ and  scale lengths $(r_d, r_g, a)=(1, 3.9,1.9)$ kpc. The bulge follows a \citet{1990ApJ...356..359H} profile. There are $1.33\times10^7$ particles, $3\times10^6$ of which are gas particles.  The initial metallicity is set to $0.1 Z_{\odot }$.

\vspace{-0.5cm}
\subsubsection{Ell}
Ell is an elliptical galaxy with halo and disc/bulge baryonic properties $(M_{\rm halo}, c)=(1.4 \times10^{13} $M$_\odot, 6)$ and $(M_{\rm bar}, M_b, M_d, M_g)=(15, 14, 1.4, 0.1)\times 10^{10} $M$_\odot$, respectively. The baryonic components have scale lengths $(r_d, r_g, a)=(4.0,4.3,3.9)$ kpc. The bulge obeys a \citet{1990ApJ...356..359H} profile. Besides the gas disc, this galaxy contains an extended live hot gas halo\footnote{Ideally, hot haloes should be included in the other simulated massive isolated galaxies (eg. HiZ and MW). However, because we focus on the ISM properties in the disc and evolve these galaxies for only a few 100 Myr, before the ISM gas is depleted, the lack of hot haloes does not significantly affect our results.} of mass $M_{\rm gas}=8.6\times10^{11}\,M_{\odot}$, initialized with a spherically-symmetric $\beta$ profile with core radius equal to the halo scale radius and $\beta=3/2$, with an initial temperature profile given by hydrostatic equilibrium and a small angular momentum corresponding to a spin parameter $\lambda=0.033$. We use $3\times10^7$ particles, $1.2\times 10^7$ of which are gas particles.   The initial metallicity is set to $ Z_{\odot } (0.05+0.95 / (1.+ (r /10\hbox{kpc})^3))$.

\vspace{-0.5cm}
\subsubsection{CosmoMW}
CosmoMW is a fully cosmological zoom-in simulation from the suite presented in \citet{2014MNRAS.445..581H}, specifically the {\bf m12i} simulation therein, chosen because it produces a galaxy with stellar mass and morphology similar to the Milky Way. 
The run uses the zoom-in method  \citep{1985Pthesis,1993ApJ...412..455K} to follow the formation history of the galaxy from an initial redshift $z>100$ to $z=0$. 
The main halo has a total mass of $\sim10^{12} $M$_{\odot}$ at $z=0$ and a typical merger and growth history for halos of its mass. We use $2.07\times10^7$ total particles ($8.82\times10^6$ gas). For this analysis, we only follow the most-massive main-progenitor halo (i.e. the center of the zoom-in region) and focus on the particles in the central region (defined as $<0.1$ virial radius).

 \vspace{-0.5cm}
\subsubsection{CosmoDwarf}
CosmoDwarf is a another cosmological zoom-in from \citet{2014MNRAS.445..581H}, specifically the {\bf m10q} simulation, chosen to be a representative dwarf galaxy -- specifically one with a $z=0$ halo mass of $\sim 10^{10}\,{\rm M}_{\odot}$ and typical merger and growth history. We use $3.87\times10^{7}$ total particles ($1.57\times10^{7}$ gas). Again we focus only on the main progenitor galaxy.
 
 
\vspace{-0.5cm}
\subsection{Cooling, star formation, and stellar feedback}

The baryonic physics of cooling, star formation, and stellar feedback follow the implementation in \cite{2014MNRAS.445..581H}. In what follows, we summarize the key aspects and focus on the new physics added for this study.

\vspace{-0.5cm}
\subsubsection{Cooling}

Cooling is followed from $10^{10}$ K to $10$K, with 11 separately tracked species followed species-by-species \citep[see e.g.][]{2009MNRAS.393...99WWW}. The low-temperature (metal fine-structure and molecular) cooling rates and ionization state are tabulated from a compilation of CLOUDY runs \citep[as in][]{2009MNRAS.393...99WWW,2008ApJ...680.1083R}, including the effects of a redshift-dependent photo-ionizing background \citep[from][]{2009ApJ...703.1416F} and local ionizing sources as described below.

\vspace{-0.5cm}
\subsubsection{Star formation}

Star formation is allowed only from gas that is locally self-gravitating (where we follow \citealt{2013MNRAS.432.2647H} to estimate the local virial criterion at each point in the simulation), is self-shielding molecular (where the molecular fraction is estimated following \citealt{2011ApJ...729...36K}), and exceeds a density $n>100\,{\rm cm^{-3}}$.\footnote{Except for the CosmoMW run with MHD, the self-gravitating criterion does not account for magnetic pressure in order to be consistent with the runs without magnetic fields. However, the magnetic field strength has to reach $100 \mu\hbox{G} \times (m_i/10^5 M_\odot)^{1/3} (n/100 \hbox{cm}^{-3})^{2/3}$ to unbind a cloud, and this is rarely the case, so it is reasonable to ignore the magnetic pressure when deciding whether a cloud is self-gravitating, at least on the relatively large scales we resolve (unlike e.g. protostellar cores).} If these criteria are met, stars form with  a rate $\dot{\rho}_\star=\rho_{\rm mol}/t_{\rm free fall}$. 
In previous studies of these star formation models, we have shown that, provided stellar feedback is explicitly included and the largest fragmentation scales in the galaxy are resolved, the galactic star formation rates and histories are regulated by stellar feedback and are insensitive to changes in these criteria (as well as more complicated chemical or temperature-based star formation models); see \citet{2011MNRAS.417..950H,2012MNRAS.421.3488H,2013MNRAS.432.2647H, 2013MNRAS.433.1970F}.  We have confirmed these studies explicitly with our ICs and simulations both including and excluding the additional microphysics we study here.

\vspace{-0.5cm}
\subsubsection{Stellar feedback}
A star particle inherits its metallicity from its parent gas particle, and is treated as a single stellar population. The feedback quantities (including luminosity, SN rates, mass and metal loss rates, etc.) are tabulated from {\sc starburst99} \citep{1999ApJS..123....3L} assuming a \cite{2002Sci...295...82K} IMF. 
Our stellar feedback model includes the following processes: (1) an approximate treatment of local and long-range momentum deposition from radiation pressure, including both initial single-scattering of optical/UV photons and (potentially) multiple-scattering of IR photons; (2) SNe (Types Ia and II), which occur stochastically according to the tabulated rates and, when they occur, deposit the appropriate ejecta energy, momentum, mass, and metals into the surrounding  gas particles; (3) stellar winds from O-stars and AGB stars, which are treated similarly to SNe except that the injection is continuous; (4) photo-ionization and photo-electric heating, with each star particle acting as a source and the UV flux incident on a gas particle estimated by accounting for self-shielding and absorption from intervening material.

\section{Additional physics}\label{s:add_ph}

\subsection{Magnetic fields (MHD)}

We treat magnetic fields in the ideal-MHD limit, using the {\small GIZMO} implementation in the MFM mode described in \cite{2016MNRAS.455...51H}. The tests described in \cite{2016MNRAS.455...51H} show that this MHD implementation correctly captures traditionally difficult phenomena such as the growth rates of the magneto-rotational instability (MRI), magnetic jet launching by discs, and magnetized fluid mixing (RT and KH) instabilities. Compared to SPH MHD methods (e.g. the P-SPH MHD mode in {\small GIZMO}; \citealt{2016MNRAS.455...51H}), this method is generally significantly more accurate, exhibits better convergence properties, and requires no artificial viscosity or use of an extremely large kernel size to suppress errors. For the tests presented in \citet{2016MNRAS.455...51H}, state-of-the-art grid codes (e.g., ATHENA; \citealt{2008ApJS..178..137S}) can converge to the correct solution with a similar level of accuracy. However, the method employed here typically converges to a desired accuracy more quickly, specifically in problems in which advection, angular momentum conservation, self-gravity and/or following highly compressive flows are important (problems where Lagrangian methods have advantages).
 
When magnetic fields are present, the homogenous Euler equations of hydrodynamics are replaced by their MHD versions. In a reference frame with velocity $\mathbf{v}_{frame}$, they can be written as a set of hyperbolic PDEs of the form 
\begin{equation}
\frac{\partial \mathbf{U}}{\partial t}+\nabla\cdot(\mathbf{F}-\mathbf{v}_{frame}\otimes \mathbf{U})=\mathbf{S},
\label{Eq:continuity}
\end{equation}
where $\mathbf{U}$ is the state vector of the conserved quantities, $\mathbf{F}$ is the flux vector of the conserved variables, and $\mathbf{S}$ is the source vector. 
In the pure MHD case, $\mathbf{U}$ and $\mathbf{F}$ can be written in the form 
\begin{eqnarray}
&\mathbf{U}=\left(
\begin{array}{c}
\rho  \\
\rho \mathbf{v} \\
\rho e\\
\mathbf{B}
\end{array} \right)\nonumber\\
&\mathbf{F}=\left(
\begin{array}{c}
\mathbf{F}_\rho \\
\mathbf{F}_P \\
\mathbf{F}_e\\
\mathbf{F}_B
\end{array} \right)
:=\left(
\begin{array}{c}
\rho \mathbf{v} \\
\rho \mathbf{v}\otimes\mathbf{v}+P_T \mathcal{I}-\mathbf{B}\otimes\mathbf{B} \\
(\rho e+P_T)\mathbf{v}-(\mathbf{v}\cdot\mathbf{B})\mathbf{B}\\
\mathbf{v}\otimes\mathbf{B}-\mathbf{B}\otimes\mathbf{v} 
\end{array}\right),
\label{Eq:tensor}
\end{eqnarray}
where $\rho$ is the density, $e=u_{int}+|\mathbf{B}|^2/2\rho+|\mathbf{v}|^2/2$  is the total specific energy, and $P_T=P+|\mathbf{B}|^2/2$ is the total pressure\footnote{In the MHD mode, the HLLD solver is adopted. The HLLC solver is adopted otherwise.}.

To clean the non-zero $\nabla \cdot B$ resulting from numerical errors, a combination of the \cite{Dedner2002645} and \cite{1999JCoPh.154..284P} cleaning methods are applied in {\small GIZMO}, with important modifications for the Lagrangian nature of the code (see \cite{2016MNRAS.455...51H} for detail). 
In all our ICs we seed the simulation volume with a uniform initial magnetic field in the direction of the galaxy angular momentum vector. For our cosmological runs (CosmoMW and CosmoDwarf), this is a trace (sub-nG) initial field that is quickly amplified even before galaxies form. For isolated discs, we initially set $\sim 10^{-2}\,\mu$G fields, but these are quickly amplified and dominated by the field built up through a combination of the MRI, the supersonic turbulent dynamo, and the galactic fountain dynamo. For our ``Ell'' run, we also initialize the gas in the extended galactic halo with a purely azimuthal field in equipartition with the thermal energy (set to be in hydrostatic equilibrium).

\vspace{-0.5cm}
\subsection{Anisotropic conduction}

Thermal conduction is incorporated into the Euler equations as an extra diffusion term in $\mathbf{F}_e$ from \Eqref{Eq:tensor}, following the standard Spitzer-Braginskii form. This means the conduction term added to $\mathbf{F}_e$ is $\kappa\,(\hat{\mathbf{B}}\otimes\hat{\mathbf{B}})\cdot \nabla T$,  where $\hat{\mathbf{B}}$ is the unit vector along the corresponding magnetic field. $\hat{\mathbf{B}}\otimes\hat{\mathbf{B}}$ in the expression serves as a projection operator constraining the conduction energy flux to follow the magnetic field lines and makes the thermal conduction anisotropic.   

The anisotropic conduction equation is solved and consistently implemented into the MFM/MFV methods in GIZMO. 
\cite{2016arXiv160207703H} presents tests confirming that the method is numerically stable, converges with second-order accuracy (as the MHD method in {\small GIZMO} itself does), and is capable of fully anisotropic configurations (i.e.\ the conductive flux vanishes identically when $\hat{\bf B}$ and $\nabla T$ are perpendicular).

Instead of setting the conduction coefficient $\kappa$  by hand, we calculate it  self-consistently as the Spitzer conductivity \citep{1953PhRv...89..977S,1988xrec.book.....S,2003ApJ...582..162Z,2015ApJ...798...90Z,2016MNRAS.458..410K} with the form 
\begin{eqnarray}
\kappa&&=\frac{0.96k_B(k_BT)^{5/2}}{ m_e^{1/2}e^4\ln \Lambda}\frac{ F_i }{1+4.2  \ell_e /\ell_T}  \hbox{\space\space}  \notag\\
&&=\frac{4.87\times 10^{-7} F_i }{1+4.2  \ell_e /\ell_T} T^{5/2} \hbox{\space\space\space\space\space\space}  [\hbox{erg s}^{-1}\hbox{K}^{-1} \hbox{cm}^{-1} ],
\end{eqnarray}
where $F_i$ is the ionized fraction (computed self-consistently in our cooling routines), $\ln \Lambda \sim 37 $ is the Coulomb logarithm, $\ell_e\equiv 3^{3/2} (k_B T)^2/4n_e\sqrt{\pi}e^4 \ln \Lambda $ is the electron mean free path, and $\ell_T\equiv T/|\nabla T|$ is the temperature gradient scale length. The denominator accounts for saturation of $\kappa$, which occurs when electrons have large mean-free-paths (it limits the gradient scale length to the mean-free-path \citep{1977ApJ...211..135C,1988xrec.book.....S,2016MNRAS.458..410K}). The steep temperature dependence indicates that conduction is more efficient in hotter gas.

\vspace{0.5cm}
\subsection{Anisotropic viscosity}

Viscosity is incorporated into MHD through the Navier-Stokes equations, which modify the momentum flux and the energy flux in the Euler equations as
\begin{eqnarray}
&&\mathbf{F}_P=\rho \mathbf{v}\otimes\mathbf{v}+P_T \mathcal{I}-\mathbf{B}\otimes\mathbf{B} +\Pi\nonumber\\
&&\mathbf{F}_e=(\rho e+P_T)\mathbf{v}-(\mathbf{v}\cdot\mathbf{B})\mathbf{B}+\Pi\cdot\mathbf{v}.
\end{eqnarray}
For MHD, the anisotropic viscosity again follows the Spitzer-Braginskii anisotropic form, in which the viscious flux $\Pi$ is
\begin{eqnarray}
\Pi=-3\eta\left(\hat{\mathbf{B}}\otimes\hat{\mathbf{B}}-\frac{1}{3}\mathcal{I}\right)\left(\hat{\mathbf{B}}\otimes\hat{\mathbf{B}}-\frac{1}{3}\mathcal{I}\right):\nabla\mathbf{v},
\end{eqnarray}
where ``$:$'' is defined by $\mathbf{A}:\mathbf{B}\equiv $Tr$(\mathbf{A}\cdot\mathbf{B})$.  Anisotropic viscosity is also solved and consistently implemented in the MFM/MFV methods of {\small GIZMO}, with the same convergence and stability properties as anisotropic conduction.

The viscous coefficients  are calculated  self-consistently as the leading-order Braginskii  viscosity \citep{1965RvPP....1..205B,1988xrec.book.....S,2006MNRAS.371.1025S,2015ApJ...798...90Z}, where the shear viscosity coefficient is
\begin{eqnarray}
\label{eqn:viscosity}
\eta =&&0.406\frac{m_i^{1/2} (k_BT)^{5/2}  }{(Ze)^4\ln \Lambda}\frac{F_i}{1+4.2  \ell_e /\ell_T} \hbox{\space\space\space\space\space\space} \notag\\
=&&\frac{4.5\times10^{-17}F_i}{1+4.2  \ell_i /\ell_{|v|}} T^{5/2}\hbox{\space\space\space\space\space\space} [\hbox{g s}^{-1}\hbox{cm}^{-1}],
\end{eqnarray}
and the bulk viscosity vanishes. 
Here, $m_{i}$ is the average ion mass, $m_e$ is the electron mass, $\ell_i$ is the ion mean free path and $\ell_{|v|}$ is the scale length of velocity.

\vspace{-0.5cm}
\subsection{Smagorinski (unresolved sub-grid eddy) models for metal diffusion}

Metal mixing on large scales is resolved in the simulation. However, unlike other numerical methods with mass exchange,\footnote{The numerical diffusivity of MFV or {\small AREPO} \citep{2010MNRAS.401..791S} is roughly of the scale of the velocity dispersion (or sound speed) times the resolution scale, $\sim v_t \Delta x$,  as their ``partitions of volume'' move with the flow. {\small ATHENA} \citep{2008ApJS..178..137S} could have a larger diffusion owing to its use of a fixed grid, in which case the velocity term in the diffusivity can be dominated by the bulk motion.} since our code is strictly Lagrangian, mass elements (including metals) are conserved on a per-particle basis unless they are injected directly by SNe or stellar winds. This limits spurious numerical diffusion but implies that un-resolved small-scale diffusion between particles is ignored.

Sub-grid models have been proposed to model this un-resolved transport. Because the systems we are simulating generally have extremely high Reynolds numbers, the un-resolved diffusion is usually dominated by small turbulent eddies rather than e.g.\ Brownian motion. The former is commonly approximated (see e.g. \citealt{2010MNRAS.407.1581S}) following \citet{1963MWRv...91...99S} by treating the metals as a passive scalar which obey the following diffusion equation: 
\begin{eqnarray}
&&\frac{\partial \mathbf{M}_i}{\partial t}+\nabla\cdot(D \nabla\mathbf{M}_i)=0\nonumber\\
&&D=C ||\mathbf{S}||_f \mathbf{h}^2,
\end{eqnarray}
where $\mathbf{h}$ is the resolution scale (at which the sub-grid model acts; here, it is the mean inter-particle separation within the kernel function, the equivalent of the cell size $\Delta x$ in Eulerian codes) and $C$ is the Smagorinsky-Lilly constant, calibrated from direct numerical simulations. $C$ usually ranges from 0.1 to 0.2, as calculated from Kolmogorov theory\citep{1963MWRv...91...99S,lilly1967representation,2008MNRAS.387..427W,2016arXiv161006590C}, and is set to 0.15 in our simulations. Note that this coefficient was set to 0.05-0.1 in some previous works \citep{ 2010MNRAS.407.1581S,2013ApJ...765...89S,2014MNRAS.443.3809B,2016ApJ...822...91W}, and this value was shown to be sufficient to provide a level of diffusion comparable to that of grid codes \citep{2008MNRAS.387..427W}. 
$\mathbf{S}$ is the symmetric traceless shear tensor defined as 
\begin{eqnarray}
\mathbf{S}=\frac{1}{2}(\nabla \mathbf{v}+(\nabla \mathbf{v}) ^T)-\frac{1}{3} {\rm Tr}(\nabla \mathbf{v}),
\end{eqnarray}
for which the diffusion vanishes in  purely compressive or  rotating flows. The norm in the expression is the Frobenius norm.

This model for sub-grid metal diffusion  is implemented in {\small GIZMO} following \cite{2010MNRAS.407.1581S}. However, because our resolution is much higher than many of the simulations in which it has been used before, the sub-grid diffusivity is much smaller. Moreover, we stress the importance of proper calibration of the constant $C$, which can change the diffusivity by factors of $\sim100$. We also caution that, as we will show in detail in a forthcoming work, this estimator can be very noisy in SPH methods (unlike the finite-volume methods used here), owing to zeroth-order errors in the SPH gradient estimator triggering artificial diffusion. Finally, we stress that this model assumes the motion seen in ${\bf S}$ is entirely due to turbulent flows. If there is real bulk motion (e.g.\ shear in a self-gravitating disc), this estimator will be triggered artificially. Therefore the estimated turbulent diffusivity using this simplistic sub-grid model is almost certainly an over-estimate of the real turbulent diffusivity. In future work, we will present a detailed study attempting to calibrate and rescale this model for situations where the contribution from galactic rotation is important on the resolution scale (Colbrook et al., in prep.). Our preliminary work suggests this estimator may over-estimate the true diffusivity by an order of magnitude in some cases.

\vspace{-0.5cm}
\section{Results} \label{S:results}

We simulate the ICs detailed in \sref{s:ic} with four distinct combinations of physics: 
\begin{itemize}
\item{{\bf Hydro}: Stellar feedback is not included (cooling, star formation, and self-gravity are, however). No additional microphysics (MHD, conduction, viscosity, metal diffusion) are included.}
\item{{\bf Hydro+MHD}: Stellar feedback is not included, but MHD are included.  Additional diffusion microphysics (conduction, viscosity, metal diffusion) are not included.}
\item{{\bf FB}: Stellar feedback is included, but the additional microphysics (MHD, conduction, viscosity, metal diffusion) are not.}
\item{{\bf FB+MHD}: Stellar feedback and MHD are included, but additional diffusion microphysics (conduction, viscosity, metal diffusion) are not.}
\item{{\bf FB+MHD+Micro}: Stellar feedback, MHD, and the additional diffusion operators (conduction, viscosity, metal diffusion) are all included.}
\end{itemize}
We analyze these four variants for each of the ICs below, with two exceptions:  the two cosmological runs with no feedback (Hydro) are prohibitively expensive. We are able to run the simulation to $z\sim6$, where the extremely high-density objects formed via the high star formation efficiencies in the absence of feedback force exceedingly small timesteps. However, some weak-feedback variations of the CosmoMW IC and the CosmoDwarf IC are presented in \citet{2014MNRAS.445..581H}; these are consistent with all of our other conclusions in this paper regarding the role of feedback.

The Hydro+MHD mode is only run for the MW and SMC ICs in order to demonstrate that magnetic fields have a relatively small effect regardless of whether strong stellar feedback is included, and almost all of the differences between Hydro runs and FB+MHD runs are a result of feedback.

Cosmological simulations are highly nonlinear and evolved for the entire age of the Universe. Because the equations (even of gravity alone) are formally non-linear, a small perturbation can be amplified and result in surprisingly large differences at low redshift, making it nontrivial to distinguish the systemic effects of fluid microphysics from any  particular change owing to stochastic effects. In each of the following CosmoMW plots, we estimate the magnitude of stochastic effects from 5 independent CosmoMW FB runs with small variations in the SNe coupling scheme (effectively we randomly ``re-shuffle'' the fraction of the SNe energy and momentum each neighbor particle sees to  generate random perturbations to the system). These differences have minor systematic effects on the stellar mass formed but serve the intended purpose of introducing small perturbations between the calculations. 
The shaded regions in the plots indicate the regions of parameter space spanned by these 5 runs. Owing to computational constraints, we did not perform such an experiment for the CosmoDwarf runs or for the FB+MHD+Micro CosmoMW runs, but we expect that the magnitude of stochastic effects in these simulations should be of similar size to those in the CosmoMW FB simulations.

\vspace{-0.5cm}
\subsection{Star formation histories}\label{s:SFR}

\begin{figure}
\centering
\includegraphics[width=8.5cm]{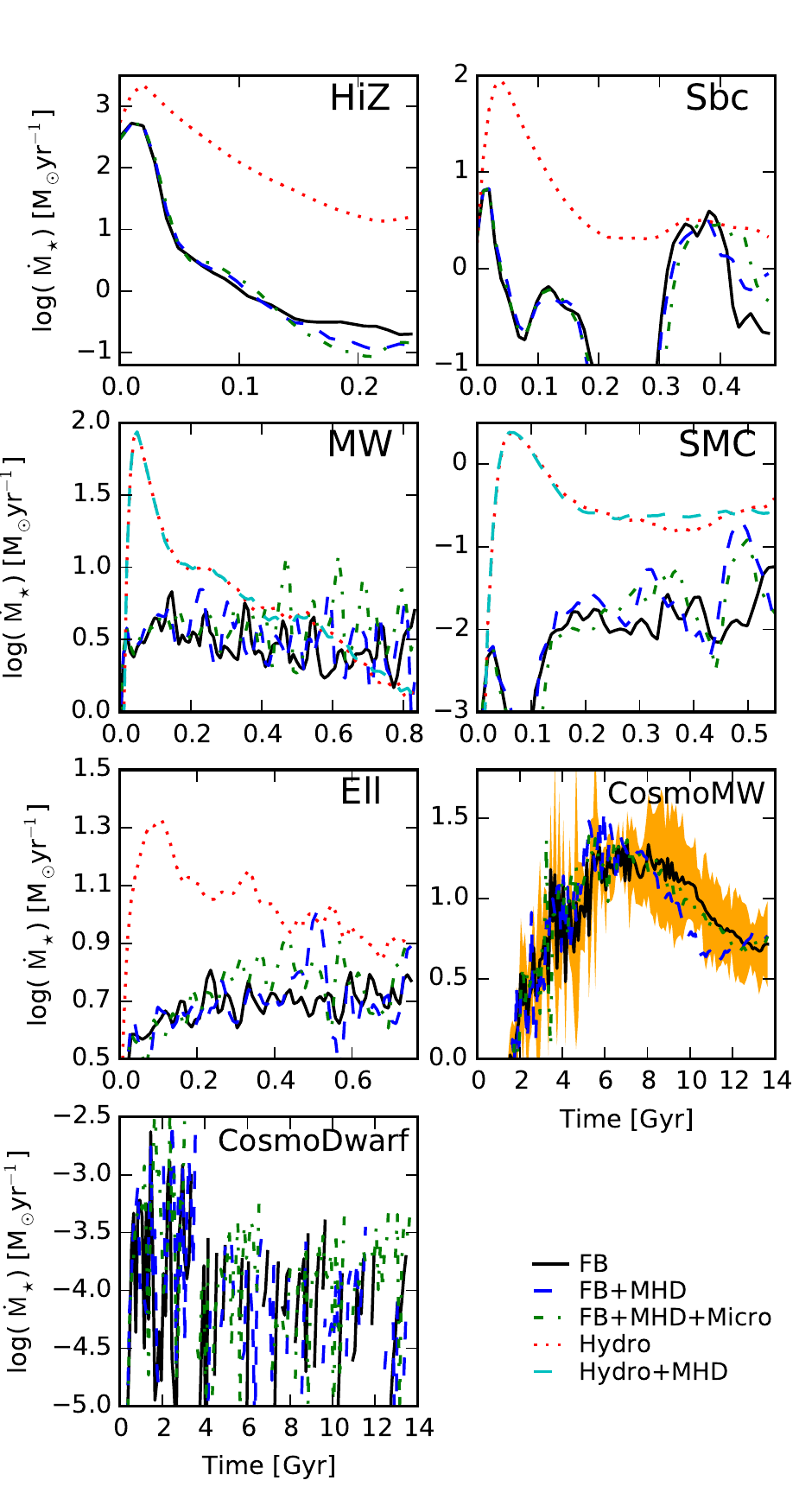}

\caption{Star formation rates (SFRs) as a function of time in each of our simulations (each IC from \tref{tab:ic}, as labeled) smoothed over 20 Myr for isolated galaxy simulations and 100 Myr for cosmological simulations so that systematic differences are clearer. HiZ (massive starburst), Sbc (dwarf starburst), MW (Milky-Way analogue), SMC (SMC-mass dwarf), and Ell (massive elliptical with a ``cooling flow'' halo) are all isolated (non-cosmological) simulations and are thus run for only a few galaxy dynamical times. Because the CosmoMW and CosmoDwarf runs are  fully cosmological zoom-in runs of a MW-mass halo and a dwarf halo, the full evolution is shown. In each, we consider four cases: default (stellar feedback, no additional microphysics, ``FB''), default+MHD (``FB+MHD''), default+MHD+anisotropic conduction and viscosity+sub-grid turbulent metal diffusion (``FB+MHD+Micro''), a run without stellar feedback (``Hydro''), and  a run without stellar feedback but with MHD (``Hydro+MHD'', only performed for the MW and SMC ICs).  In the CosmoMW case, the orange shaded region indicates the range of stochastic effects (see \S \ref{S:results}). Once feedback is included, a lower, steady-state SFR emerges; the SFR has relatively small dependence on the different microphysics considered (up to stochastic effects). However, a more steady star formation history can be observed in the CosmoDwarf FB+MHD+Micro run, resulting in a slightly higher SFR on average. }
\label{fig:sfr}
\end{figure}

\begin{figure}
\centering
\includegraphics[width=8.5cm]{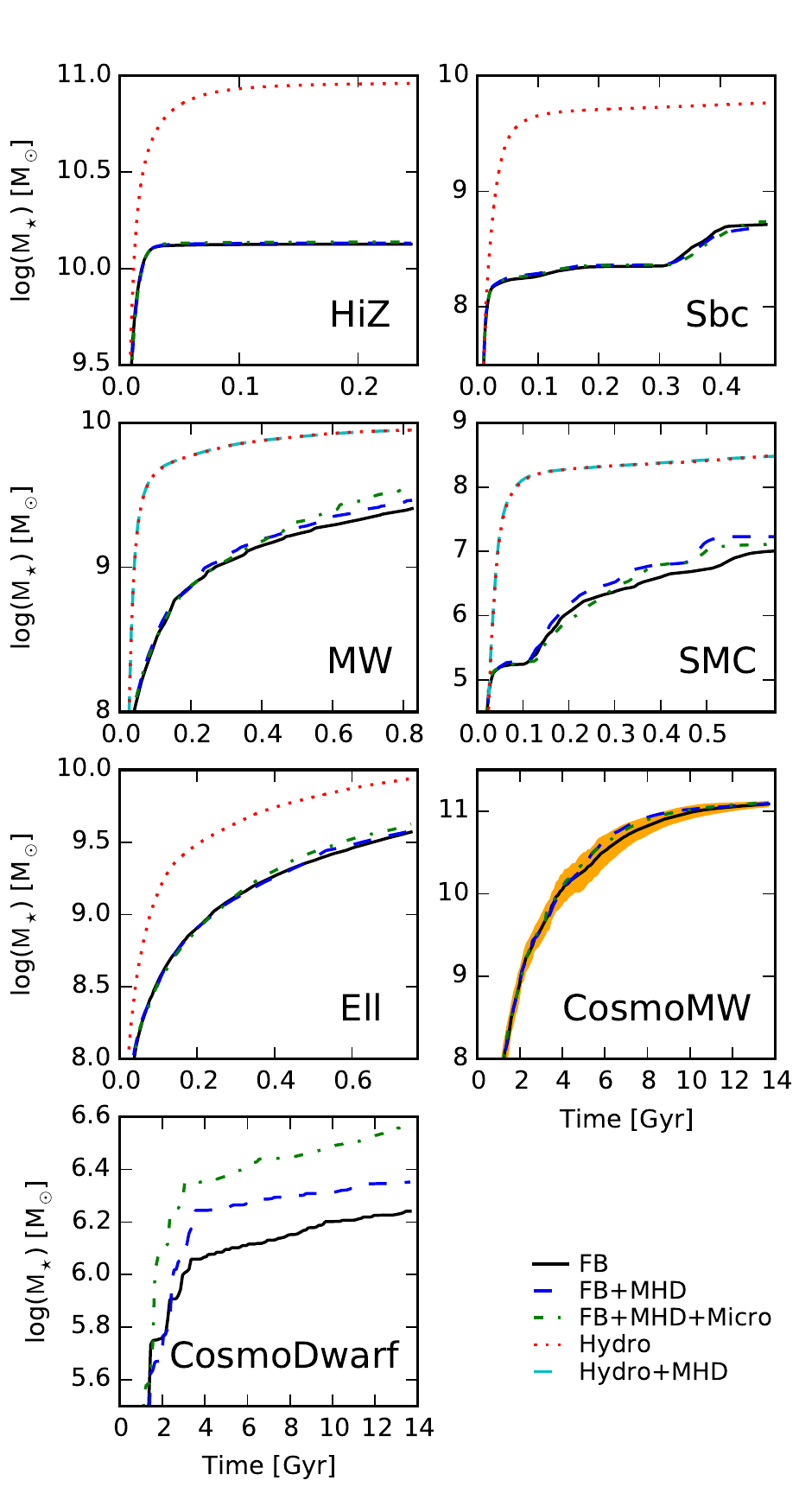}

\caption{ Total stellar mass as a function of time in each of our simulations (each IC from \tref{tab:ic}, as labeled).  The orange shaded region in CosmoMW again indicates the range of stochastic effects. Fluid microphysics has little effect on the stellar mass formed in our simulations, in contrast to stellar feedback, which reduces the stellar mass formed by about an order of magnitude relative to the no-feedback runs. However, a small hint of higher stellar mass can be observed in the runs with magnetic fields, and in the CosmoDwarf simulations, the FB+MHD+Micro run has stellar mass a factor of $\sim1.4$ higher than the FB+MHD run. These effects are generally smaller than systematic uncertainties in feedback (e.g. SNe rates). }
\label{fig:sfr_a}
\end{figure}

\fref{fig:sfr} shows the star formation histories of the five isolated galaxies, CosmoMW, and CosmoDwarf, evolved under the different combinations of physics described above. 
 The SFRs of HiZ, Sbc, MW, SMC and  Ell shown in \fref{fig:sfr} are the values averaged over 20 million years, and the SFRs of CosmoMW and CosmoDwarf are averaged over roughly 100 million years (to make systematic, as opposed to stochastic, differences clear).  To further suppress stochastic effects,  the stellar mass as a function of time is also plotted in \fref{fig:sfr_a}.

Without stellar feedback, the SFRs are generally higher than observed by factors of $\sim10-100$, regardless of whether magnetic fields are included (i.e. magnetic fields alone are insufficient to globally suppress star formation).
On the other hand, the additional physics we consider changes the SFR by $\lesssim 10\%$.  We have also confirmed that the predicted Kennicutt-Schmidt (KS)  relation is insensitive to the additional microphysics in the simulations (and in agreement with observations, as in our previous simulations with the FIRE feedback physics).

The largest effect caused by fluid microphysics can be observed in the CosmoDwarf FB+MHD+Micro run, in which the stellar mass is systemically higher than that in the FB run by a factor of $\sim1.5$. We argue below that this is most likely to be a consequence of conduction, which can increase the efficiency of cooling when multiple SNe remnants overlap (and overrun small cold gas clumps in the galaxy), thereby dissipating the hot gas energy going into galactic winds slightly more efficiently.

Compared with the other galaxies, the differences in SFRs are smallest among the different  Ell simulations (the SFR in the no-feedback run is never more than an order-of-magnitude greater than the SFRs in the variants that include explicit stellar feedback). The reason is that in Ell, the gas disc contributes  $<1\%$ of the total gas mass, while the majority is distributed in a hot gas halo. The cooling efficiency of the hot halo gas onto the star-forming disc therefore controls the gas supply available for star formation and becomes an important regulator of the SFR besides stellar feedback. 


Although magnetic fields have been suggested  in the literature as a mechanism to suppress star formation owing to their additional pressure \citep{2005ApJ...629..849P,2007ApJ...663..183P,2009ApJ...696...96W,2013MNRAS.432..176P,2012MNRAS.422.2152B}, we actually see a small hint of systemically higher stellar mass  in  the MHD runs when feedback is included. The difference is more obvious in the smaller galaxies like CosmoDwarf and SMC. But even in these cases,  the difference is less than 0.1 dex, despite the fact that magnetic field strengths consistent with observations are self-consistently obtained. On the other hand, without feedback, the star formation history is not significantly altered when magnetic fields are included.   We discuss these points further below in \sref{s:discussion}. We note, however, that magnetic fields may still play a more important role in regulating the formation of individual stars (which is unresolved in our simulations).

\vspace{-0.5cm}
\subsection{Morphologies}\label{s:morphology}

\begin{figure*}
\begin{flushleft}

\begin{minipage}[b]{0.247\linewidth}
\includegraphics[height=4.44cm]{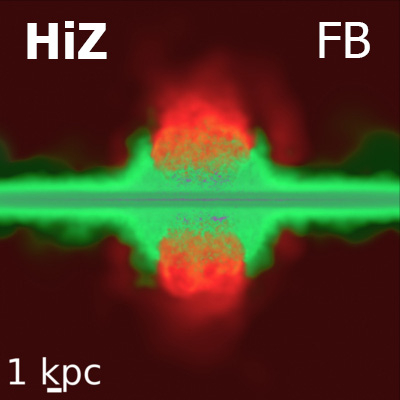}
\end{minipage}
\begin{minipage}[b]{0.247\linewidth}
\centering
\includegraphics[height= 4.44cm]{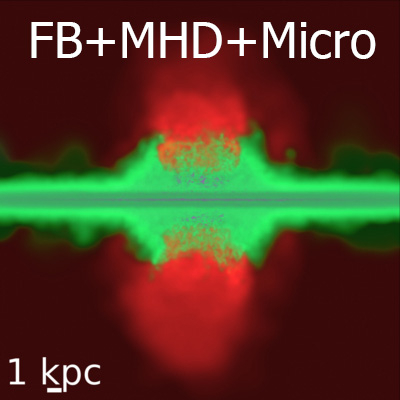}
\end{minipage}
\begin{minipage}[b]{0.247\linewidth}
\centering
\includegraphics[height= 4.44cm]{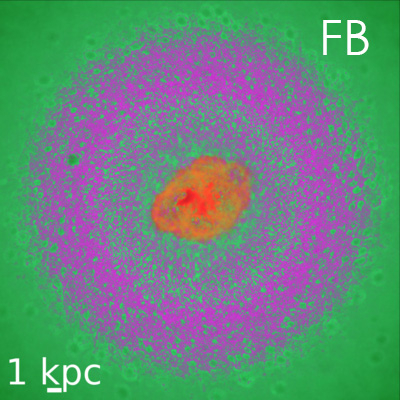}
\end{minipage}
\begin{minipage}[b]{0.247\linewidth}
\centering
\includegraphics[height= 4.44cm]{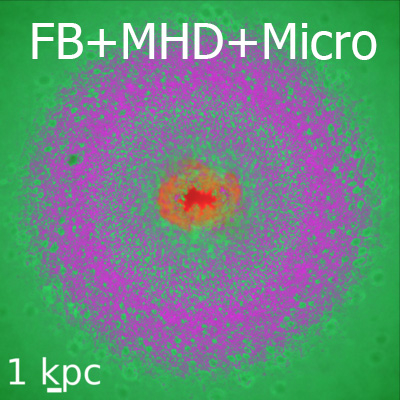}
\end{minipage}

\begin{minipage}[b]{0.247\linewidth}
\centering
\includegraphics[height= 4.44cm]{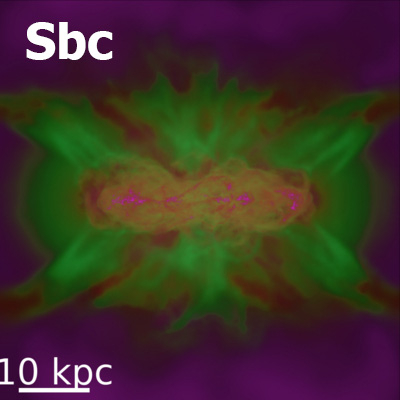}
\end{minipage}
\begin{minipage}[b]{0.247\linewidth}
\centering
\includegraphics[height= 4.44cm]{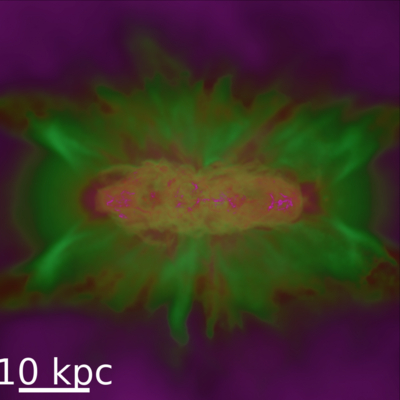}
\end{minipage}
\begin{minipage}[b]{0.247\linewidth}
\centering
\includegraphics[height= 4.44cm]{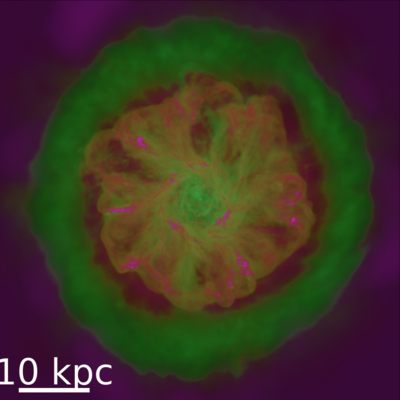}
\end{minipage}
\begin{minipage}[b]{0.247\linewidth}
\centering
\includegraphics[height= 4.44cm]{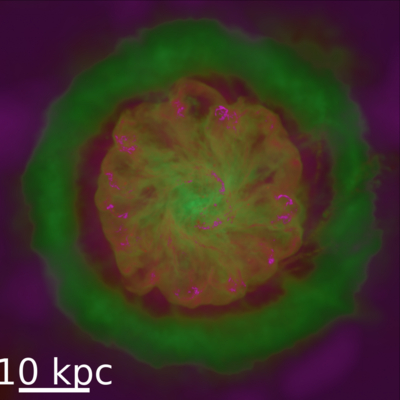}
\end{minipage}

\begin{minipage}[b]{0.247\linewidth}
\centering
\includegraphics[height= 4.44cm]{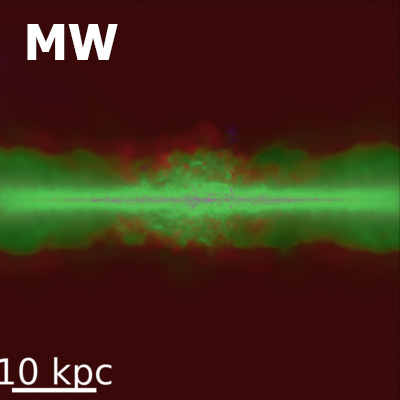}
\end{minipage}
\begin{minipage}[b]{0.247\linewidth}
\centering
\includegraphics[height= 4.44cm]{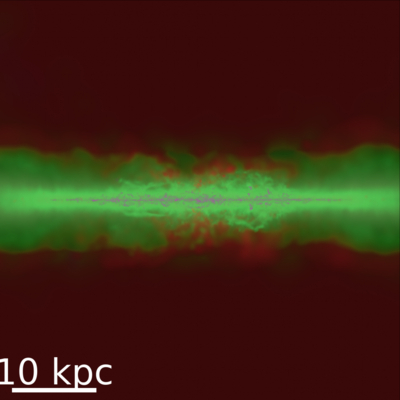}
\end{minipage}
\begin{minipage}[b]{0.247\linewidth}
\centering
\includegraphics[height= 4.44cm]{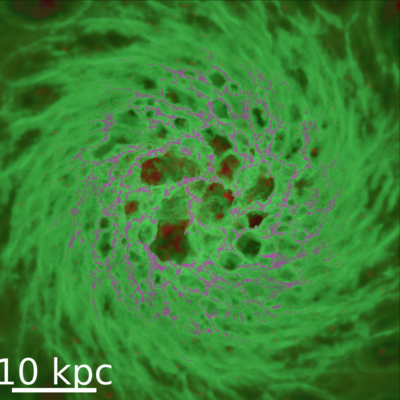}
\end{minipage}
\begin{minipage}[b]{0.247\linewidth}
\centering
\includegraphics[height= 4.44cm]{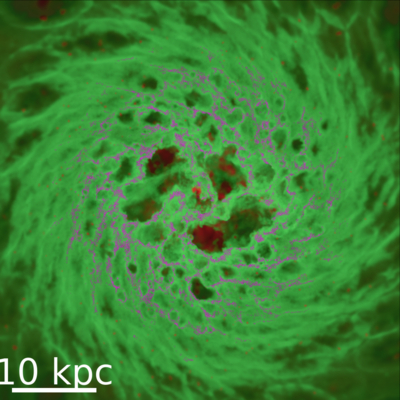}
\end{minipage}

\begin{minipage}[b]{0.247\linewidth}
\centering
\includegraphics[height= 4.44cm]{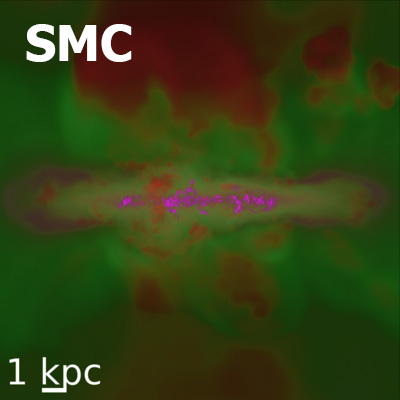}
\end{minipage}
\begin{minipage}[b]{0.247\linewidth}
\centering
\includegraphics[height= 4.44cm]{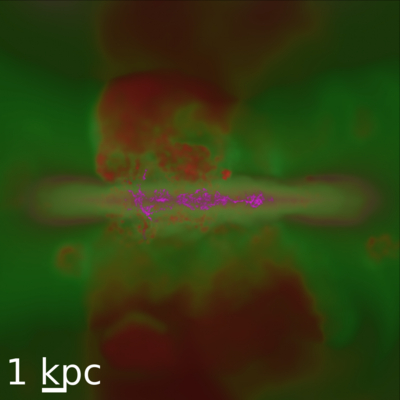}
\end{minipage}
\begin{minipage}[b]{0.247\linewidth}
\centering
\includegraphics[height= 4.44cm]{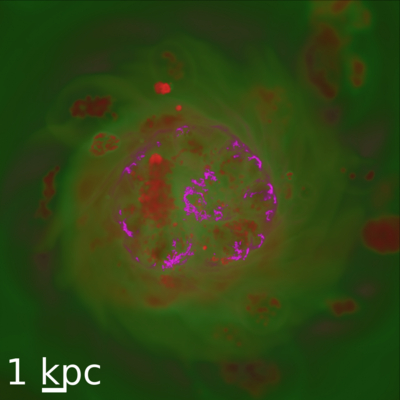}
\end{minipage}
\begin{minipage}[b]{0.247\linewidth}
\centering
\includegraphics[height= 4.44cm]{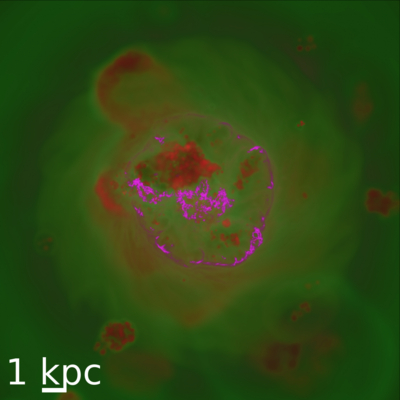}
\end{minipage}
 
\begin{minipage}[b]{0.247\linewidth}
\centering
\includegraphics[height= 4.44cm]{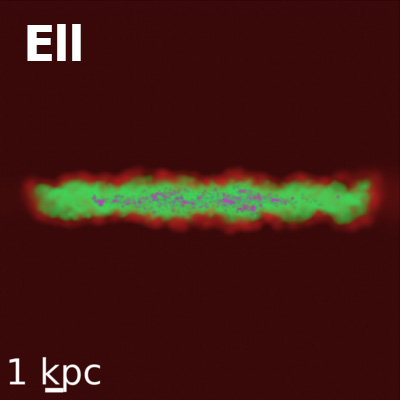}
\end{minipage}
\begin{minipage}[b]{0.247\linewidth}
\centering
\includegraphics[height= 4.44cm]{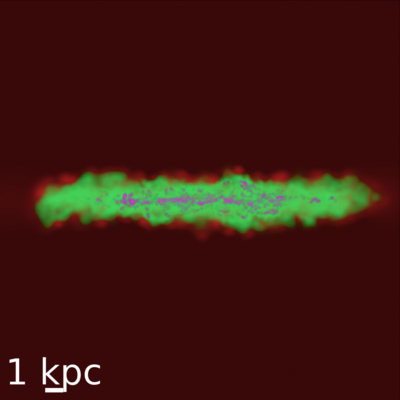}
\end{minipage}
\begin{minipage}[b]{0.247\linewidth}
\centering
\includegraphics[height= 4.44cm]{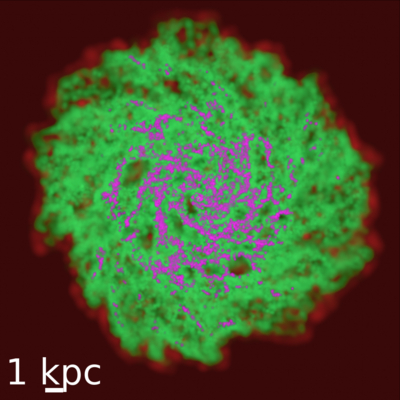}
\end{minipage}
\begin{minipage}[b]{0.247\linewidth}
\centering
\includegraphics[height= 4.44cm]{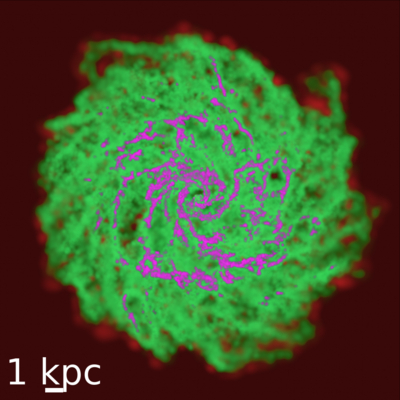}
\end{minipage}
\label{fig:morph}
\end{flushleft}
\caption{Images of the gas morphology of the isolated galaxies with feedback. The intensity encodes the projected density (log-weighted with $\sim4$ dex stretch); different temperatures are shown in red ($>10^{5}$\,K), green ($8000 - 10^5$\,K), and magenta ($<8000\,$K). We show edge-on and face-on projections for our FB and FB+MHD+Micro (FB+MHD is similar).  The morphologies of the runs with the same stellar feedback but different additional physics show little difference. 
}
\end{figure*}

\begin{figure*}
\begin{flushleft}

\begin{minipage}[b]{0.247\linewidth}
\centering
\includegraphics[height= 4.44cm]{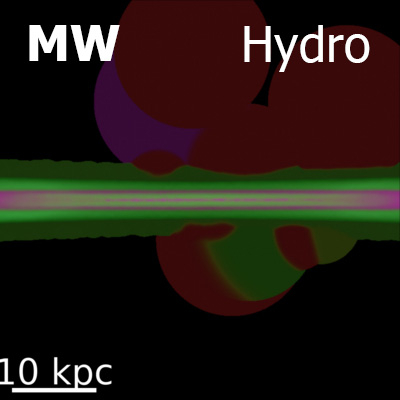}
\end{minipage}
\begin{minipage}[b]{0.247\linewidth}
\centering
\includegraphics[height= 4.44cm]{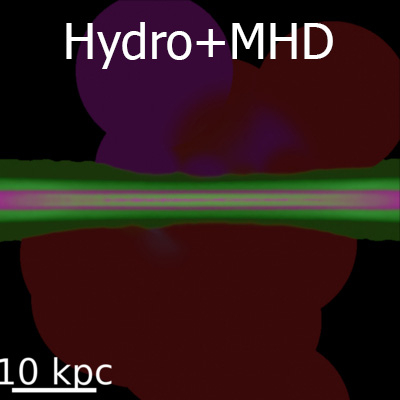}
\end{minipage}
\begin{minipage}[b]{0.247\linewidth}
\centering
\includegraphics[height= 4.44cm]{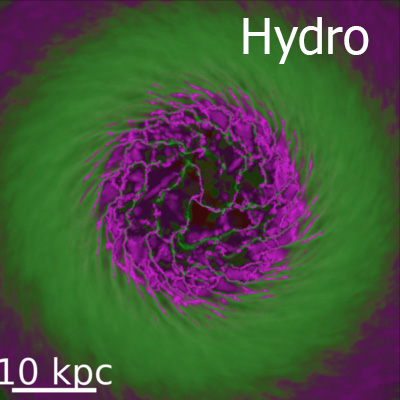}
\end{minipage}
\begin{minipage}[b]{0.247\linewidth}
\centering
\includegraphics[height= 4.44cm]{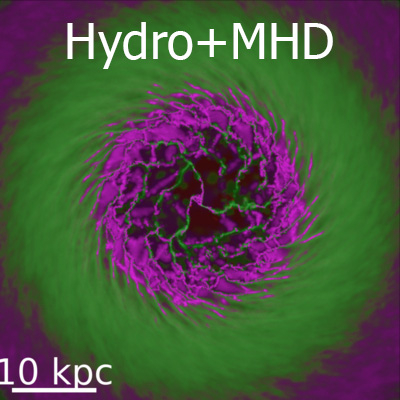}
\end{minipage}

\begin{minipage}[b]{0.247\linewidth}
\centering
\includegraphics[height= 4.44cm]{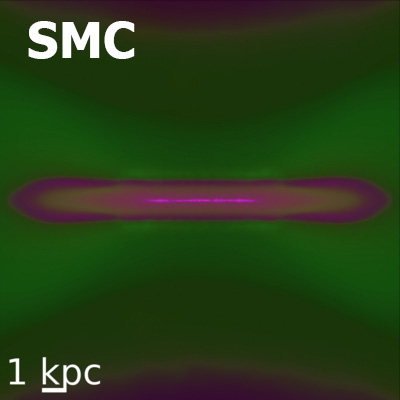}
\end{minipage}
\begin{minipage}[b]{0.247\linewidth}
\centering
\includegraphics[height= 4.44cm]{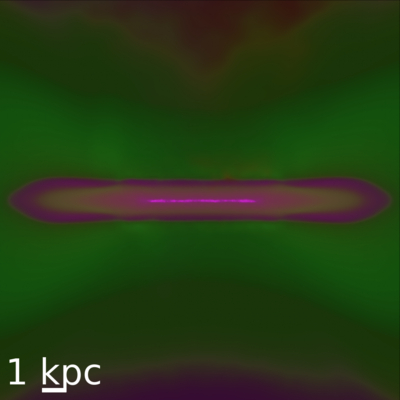}
\end{minipage}
\begin{minipage}[b]{0.247\linewidth}
\centering
\includegraphics[height= 4.44cm]{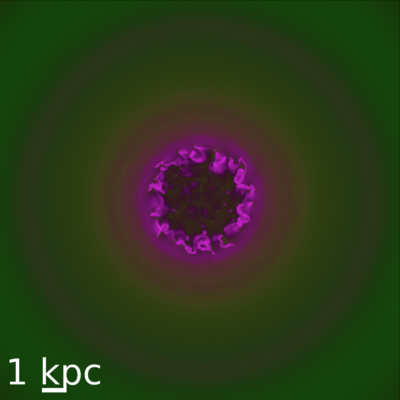}
\end{minipage}
\begin{minipage}[b]{0.247\linewidth}
\centering
\includegraphics[height= 4.44cm]{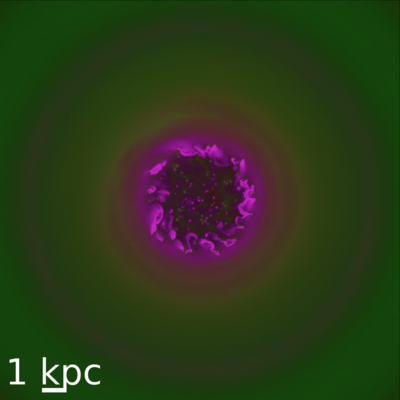}
\end{minipage}

\label{fig:morph_nofb}
\end{flushleft}
\caption{The effect of magnetic fields in the absence of feedback on the MW and SMC gas morphology. The Hydro+MHD runs exhibit morphologies that are almost identical to those of the Hydro runs. In both cases, it is clear that the inner gas discs have catastrophically fragmented and been converted into stars, thus indicating that the magnetic pressure alone is insufficient to prevent the gas from undergoing run-away collapse.
}
\end{figure*}

\begin{figure}
\begin{flushleft}

\begin{minipage}[b]{0.489\linewidth}
\centering
\includegraphics[height=4.18cm]{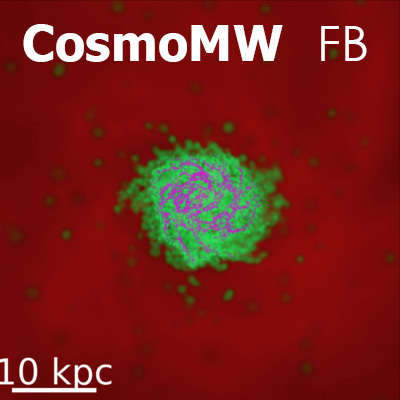}
\end{minipage}
\begin{minipage}[b]{0.489\linewidth}
\centering
\includegraphics[height=4.18cm]{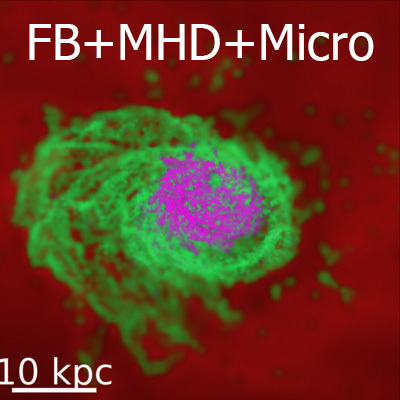}
\end{minipage}

\begin{minipage}[b]{0.489\linewidth}
\centering
\includegraphics[height=4.18cm]{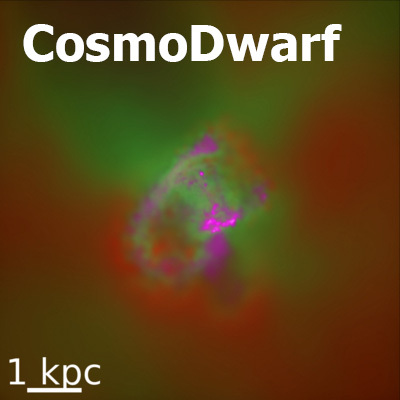}
\end{minipage}
\begin{minipage}[b]{0.489\linewidth}
\centering
\includegraphics[height=4.18cm]{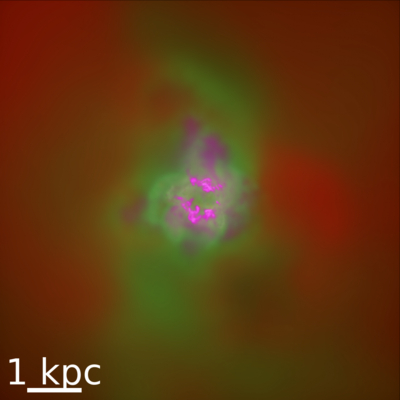}
\end{minipage}

\label{fig:morph_cos}
\end{flushleft}
\caption{Images of the gas morphology of the cosmological simulations at $z=0$, as \fref{fig:morph}. The slightly puffier disc in the CosmoMW FB+MHD+Micro run suggests a slightly higher accretion rate in the outer disc of this simulated galaxy. This owes to metal diffusion and conduction enhancing cooling from the ``hot'' circumgalactic medium (CGM).
}
\end{figure}

\fref{fig:morph}, \fref{fig:morph_nofb} and \fref{fig:morph_cos} shows the face-on and edge-on gas morphologies of our simulated galaxies after some dynamical evolution; the colours denote gas in different temperature bins (see caption).  The CosmoMW runs and CosmoDwarf runs are shown at $z\sim 0$.   The no-feedback runs in \fref{fig:morph_nofb}  (both with and without MHD) again show fundamental differences from the other runs (which incorporate explicit stellar feedback) owing to the runaway collapse of gas, as described in \cite{2012MNRAS.421.3488H,2014MNRAS.445..581H}. The magnetic pressure is unable to stop this process, and therefore the morphologies in the Hydro and Hydro+MHD runs appear essentially identical.


When stellar feedback is included, there is also no significant systematic difference in morphology among runs with different additional physics. Stochastic SNe events can make some parts of some of the variants hotter at the times shown in the figure, but there is little systematic difference in a time-averaged sense. The CosmoMW FB+MHD+Micro run seems to have a slightly more extended disc at $z\sim0$, suggesting a slightly higher accretion rate owing to turbulent metal diffusion and conduction enhancing cooling from the CGM. 
Among the different galaxies, Ell stands out as an exception for having little variation even between the runs with and without feedback. This is because cooling from the hot halo gas plays an important role in regulating Ell, as discussed above. 


\vspace{-0.5cm}
\subsection{Gas phase structure}\label{s:phase}

\begin{figure*}
\begin{flushleft}
\centering
\includegraphics[width=18cm]{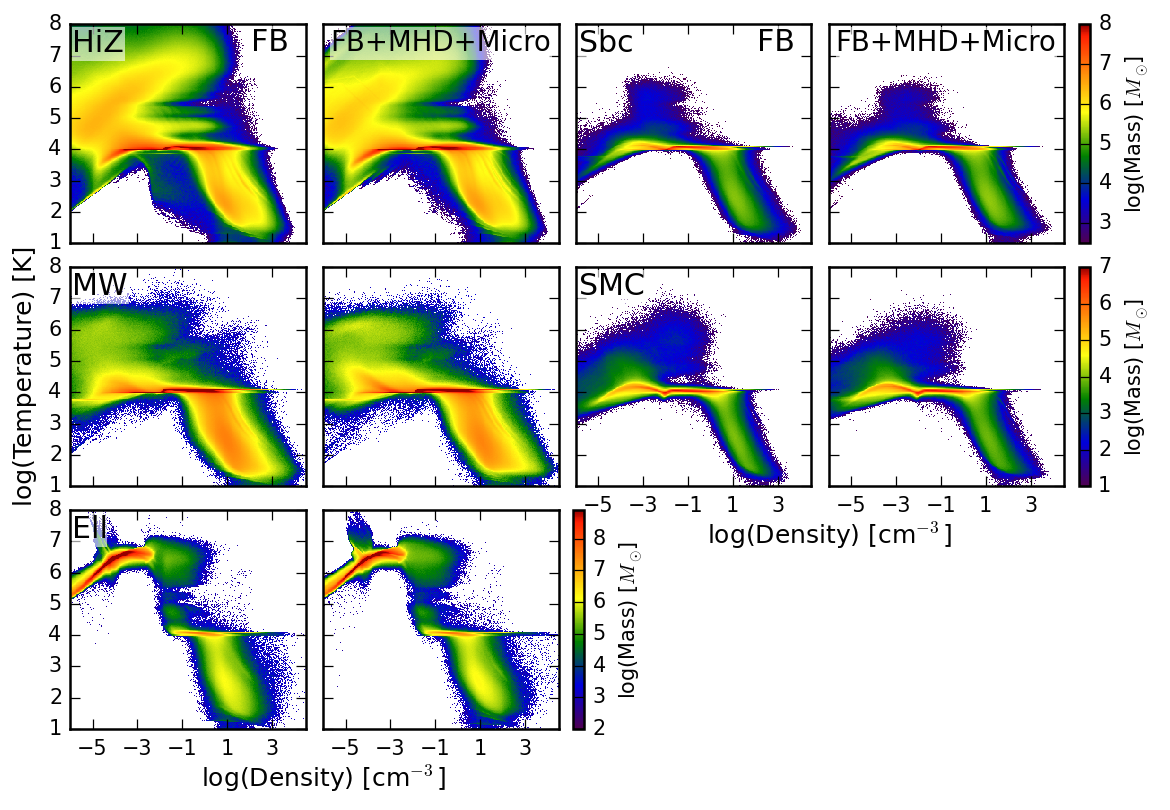}
\end{flushleft}
\label{fig:phase}
\caption{Temperature-density phase distribution of our isolated galaxy simulations. Each plot is averaged over the entire simulation duration. With feedback, cold neutral, warm ionized, and hot (feedback-driven) volume-filling phases are present. The additional MHD and diffusion microphysics have little effect on the phase structure in the presence of feedback. Note that the ``spike'' in the upper-left corners (hot halo phase) of the Ell runs correspond to an artificial shock arising from the IC being out of equilibrium.  }
\end{figure*}

\begin{figure*}
\begin{flushleft}
\centering
\includegraphics[width=18cm]{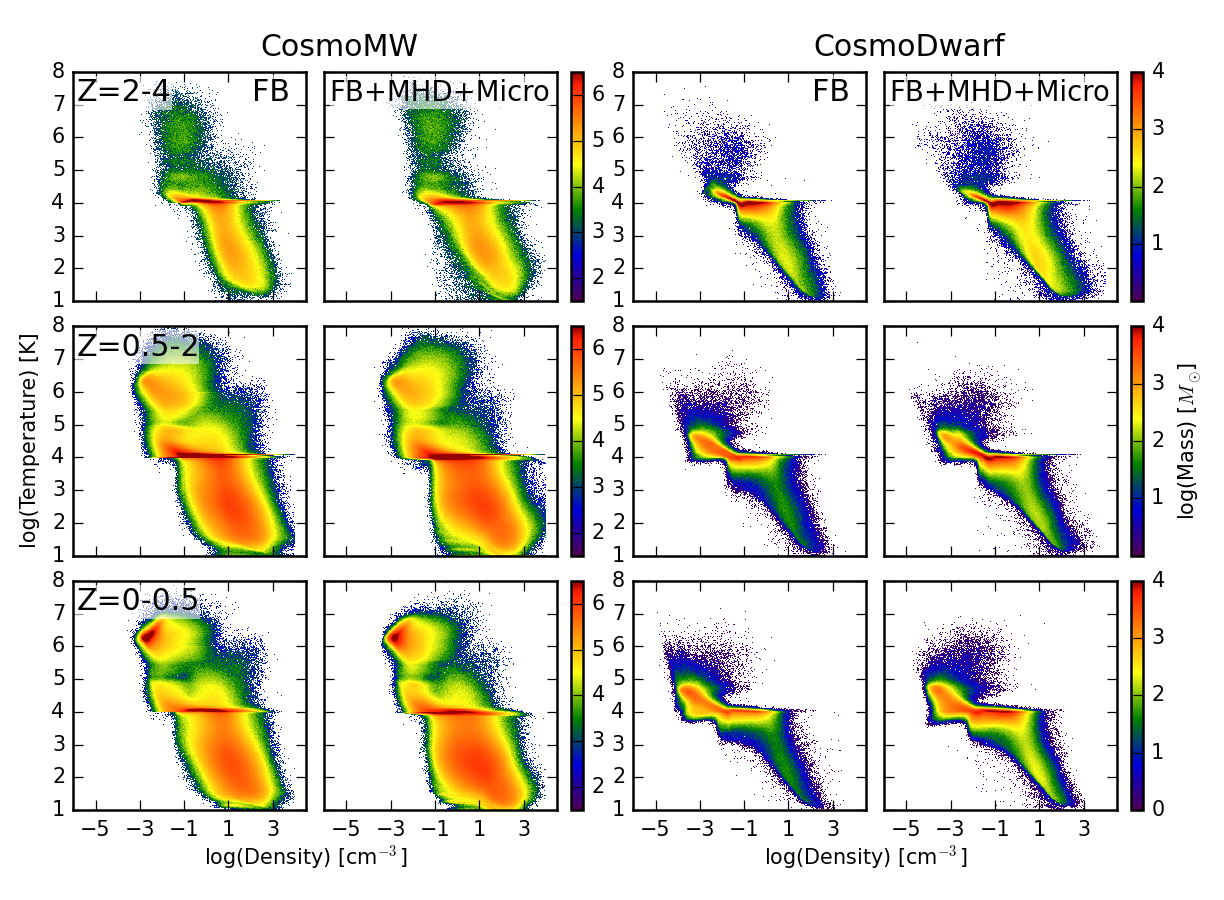}
\end{flushleft}
\label{fig:phase_cos}
\caption{Temperature-density phase distribution of the CosmoMW  and CosmoDwarf cosmological simulation. Each plot is averaged in three separate redshift intervals (labeled). Recall, only particles within $\lesssim 0.1 R_{\rm vir}$ the central galaxy are plotted, to focus on ISM properties as in \fref{fig:phase}. The additional MHD and diffusion microphysics have relatively weaker effects in the presence of feedback, although some changes in the warm, inner CGM gas ($T\sim 10^5- 10^7$K, $n>10^{-2}$ cm$^{-3}$)  are evident.}
\end{figure*}

\begin{figure*}
\begin{flushleft}
\centering
\includegraphics[width=18cm]{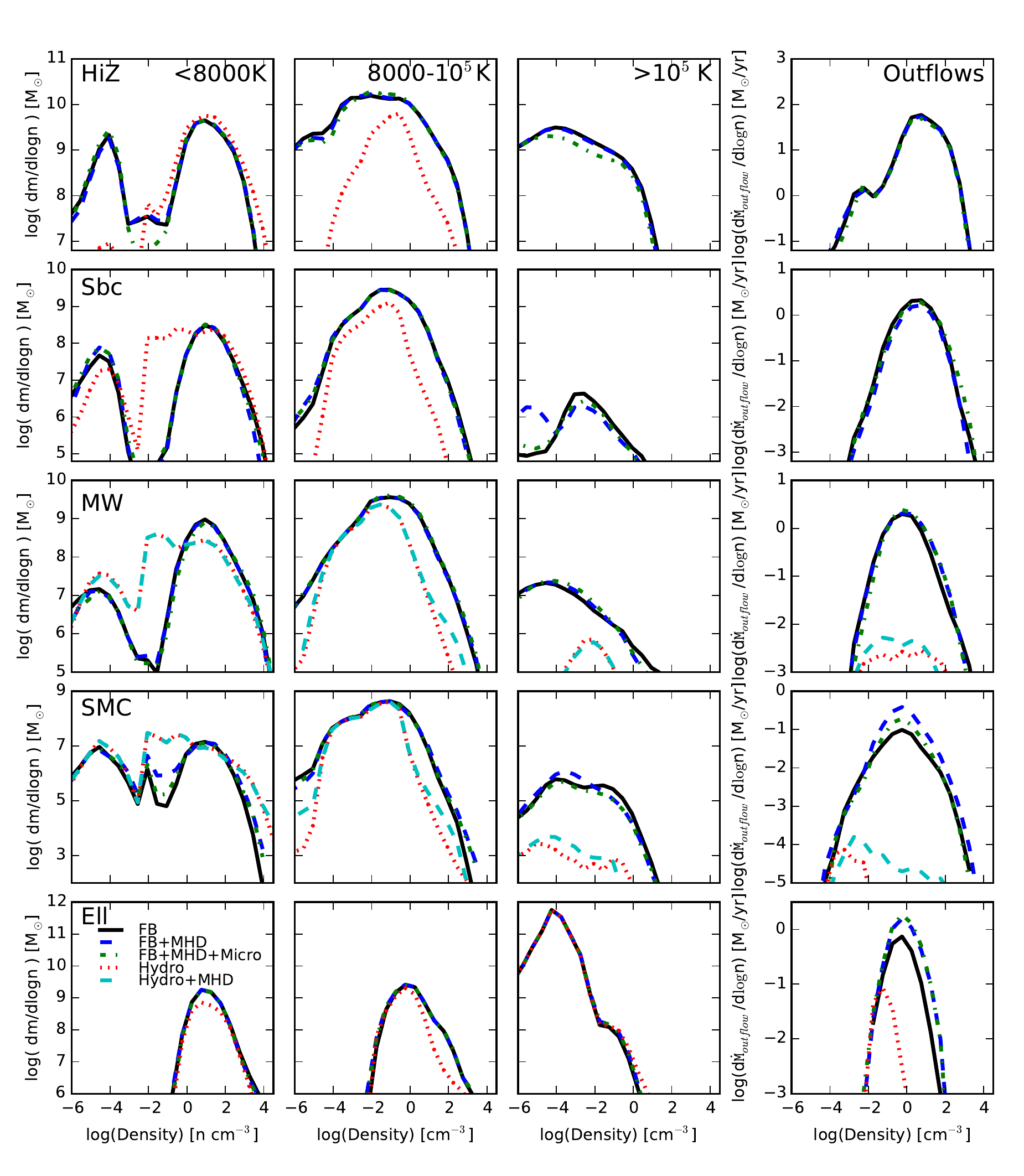}
\end{flushleft}
\label{fig:temperature}
\caption{Density distribution of gas in different phases. Rows show our isolated galaxy simulations; columns show phases including cold neutral ({\em left}), warm ionized ({\em middle left}), hot ({\em middle right}), and outflow ({\em right}). To estimate the outflow, we simply take all gas that is within 0.5 kpc of the boundary of the disc (taken as a cylinder with radius $10\,$kpc and height $2\,$kpc) and moving with a radial velocity greater than some $v_{\rm min}$ chosen to be an appreciable fraction of the escape velocity in each galaxy ($v_{\rm min}=(200,\,100,\,100,\,30,\,100)\,{\rm km\,s^{-1}}$ for HiZ, Sbc, MW, SMC, and Ell  runs, respectively). The no-feedback runs produce far less  hot gas and more cold gas, as expected (the difference is again less visible in the Ell run since the hot phase is dominated by the initial hot halo gas) with or without magnetic fields. Varying the additional microphysics has relatively weak effects. An increase in the outflow rate is visible in the run that includes magnetic fields but no feedback (Hydro+MHD), but the difference is orders of magnitude less than what is caused by feedback.  }
\end{figure*}

\begin{figure*}
\begin{flushleft}
\centering
\includegraphics[width=18cm]{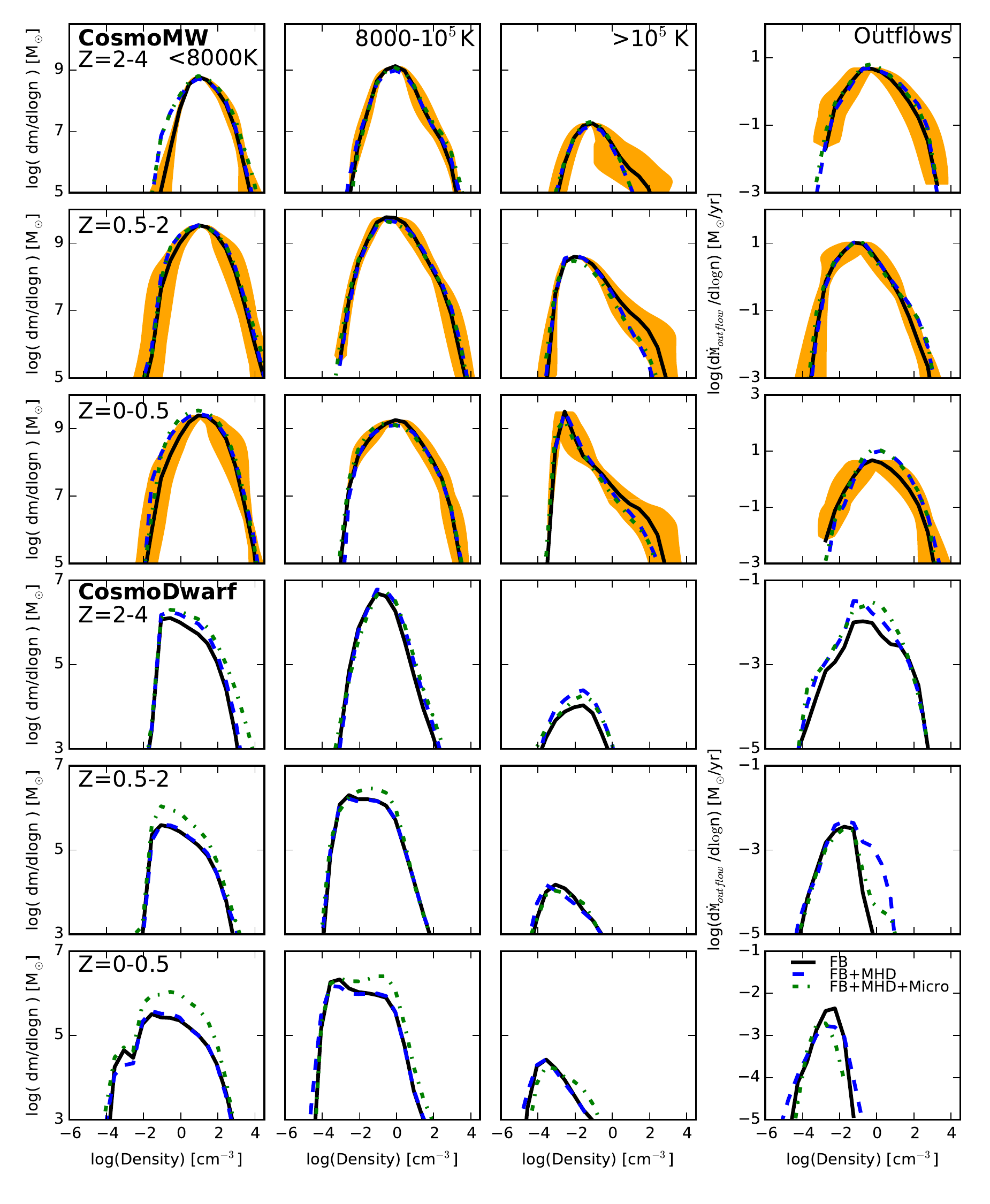}

\end{flushleft}

\label{fig:temperature_cos}
\caption{Gas density distribution in different phases, as in \fref{fig:temperature}, but for our CosmoMW and CosmoDwarf runs, averaged in different redshift intervals. In the non-outflow panels, we consider only gas within $\lesssim 0.1 R_{\rm vir}$ of the central galaxy as \fref{fig:phase_cos}. To quantify the outflows at $z = 0.5-4$ for CosmoMW and at all redshifts for CosmoDwarf, instead of using a disc, we select all gas located between $0.08$ and $0.1$ virial radii of the halo center that is moving with a radial velocity greater than $100\, {\rm  km\,s^{-1}}$.  The orange shaded regions in each panel of CosmoMW indicate the magnitude of stochastic effects (see text for details). At each redshift in both galaxies, the properties are similar in all runs with stellar feedback. 
}
\end{figure*}

\fref{fig:phase} and \fref{fig:phase_cos} compare the temperature-density phase plots of our isolated and cosmological simulations, respectively. 
The gas mass in each phase is further quantified in \fref{fig:temperature} and \fref{fig:temperature_cos}, where the density distributions in the following temperature intervals are plotted: cold neutral ($0-8000$\,K), warm ionized ($8000-10^5$\,K) and hot ($>10^5$\,K). 

With stellar feedback, a multi-phase ISM is established, with star-forming cold atomic/molecular gas, warm ionized gas, and volume-filling hot gas, which is  extensively discussed in \cite{2011MNRAS.417..950H,2012MNRAS.421.3488H,2012MNRAS.421.3522H,2013MNRAS.433...69H}. Turing off stellar feedback, on the other hand, leads to the same results as discussed in \sref{s:morphology}.

Runs with the standard stellar feedback but different additional physics are very similar. Although magnetic fields, viscosity and conduction can in principle alter the cooling efficiency and fluid mixing, these additional physics have less than a $\sim10\%$ effect on the balance of ISM phases in our simulations. In fact,  as demonstrated in the MW and SMC case of \fref{fig:temperature}, the  effect of magnetic field remain weak without stellar feedback.

The phase structure difference among the Ell runs is again small, as it is dominated by the hot halo phase, and the supply of gas to the other phases relies on and is therefore regulated by the cooling flow.

There is a small difference in our CosmoDwarf runs, where the cold gas mass is larger in our FB+MHD+Micro run by $\sim 0.2$ dex ($\sim 50 \%$)  This is consistent with the slightly higher SFR in that run.

\begin{figure}
\begin{flushleft}
\centering
\includegraphics[width=9cm]{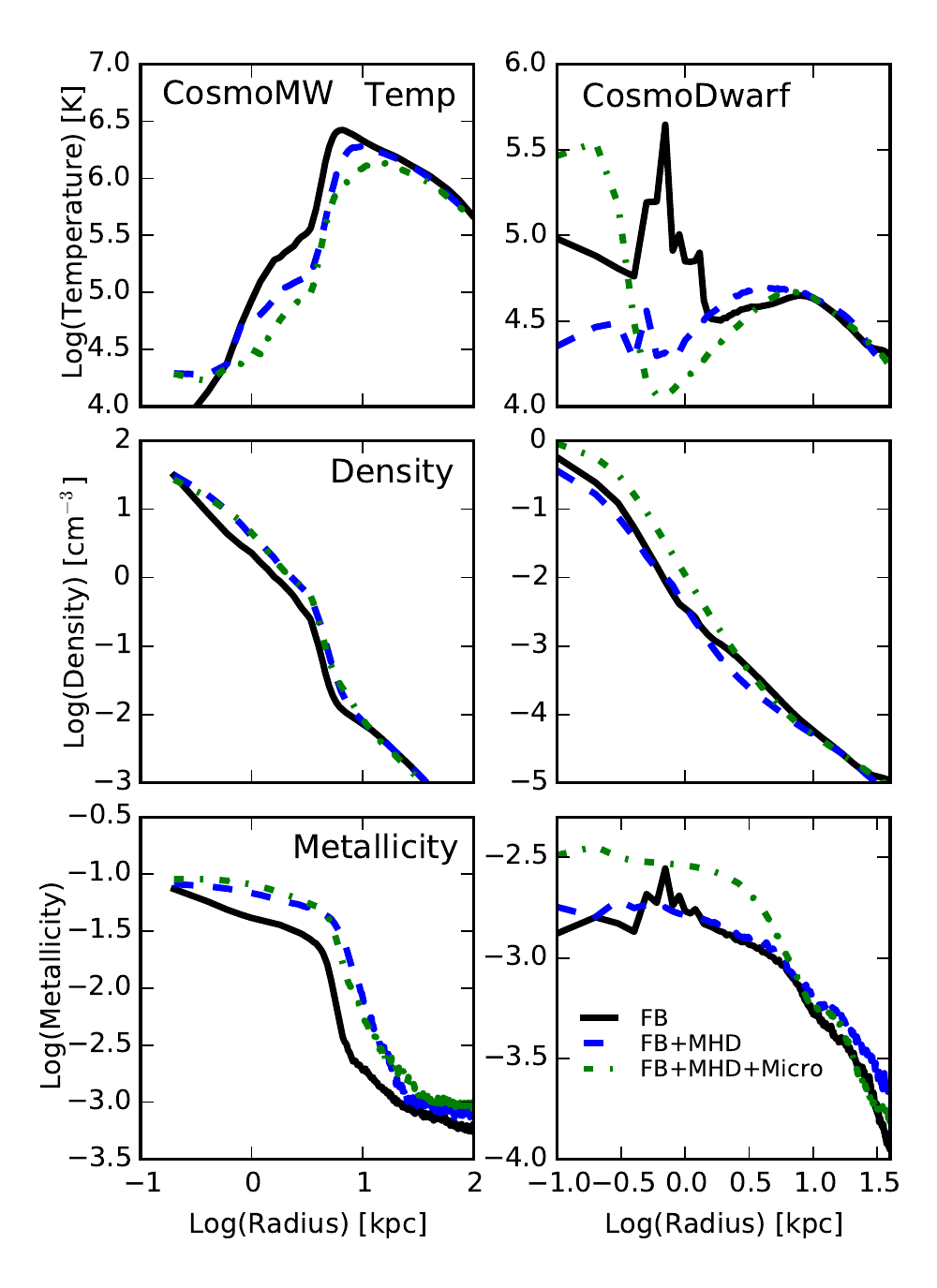}
\end{flushleft}
\label{fig:rad}
\caption{Radial distributions of temperature, gas density and metallicity for the cosmological runs averaged over the redshift range $z\sim 0-0.07$. The profiles of the CosmoMW runs
are broadly similar, with the density and metallicity (temperature) slightly higher (lower) in the runs with MHD. The temperature profiles of the CosmoDwarf runs differ significantly,
likely because this galaxy is still highly stochastic, with strong starbursts and outflows, even at $z \sim 0$. The gas density profiles are similar. The metallicity of the FB+MHD+Micro
run is higher within the central kpc, likely because of turbulent metal diffusion and the slightly higher stellar mass in this run. }
\end{figure}

\fref{fig:rad} presents the radial distributions of temperature, gas density and metallicity of the cosmological runs. The results are averaged over the redshift range $z\sim 0-0.07$ to suppress stochastic effects. For the CosmoMW runs, the profiles are broadly similar, with the density and metallicity (temperature) slightly higher (lower) in the runs with MHD. These are in consistent with the slightly more extended discs in CosmoMW FB+MHD and FB+MHD+Micro runs at low redshift as discussed in  \sref{s:morphology}. For the CosmoDwarf run, there are
marked differences in the temperature profiles; this is likely because simulated dwarfs are highly stochastic, with strong starbursts and outflows even at $z \sim 0$ \citep{2015MNRAS.454.2691M,2015arXiv151005650H}.
The gas density profiles are similar. In the FB+MHD+Micro run, the metallicity is systematically higher within the central kpc, likely because of the effects of turbulent metal diffusion and the slightly
higher stellar mass in this run; this subject will be analyzed in detail in a future work (Escala et al., in preparation).




\vspace{-0.5cm}
\subsection{Magnetic and turbulent energies}
\label{s:magnetic}

\begin{figure}
\centering
\includegraphics[width=8.5cm]{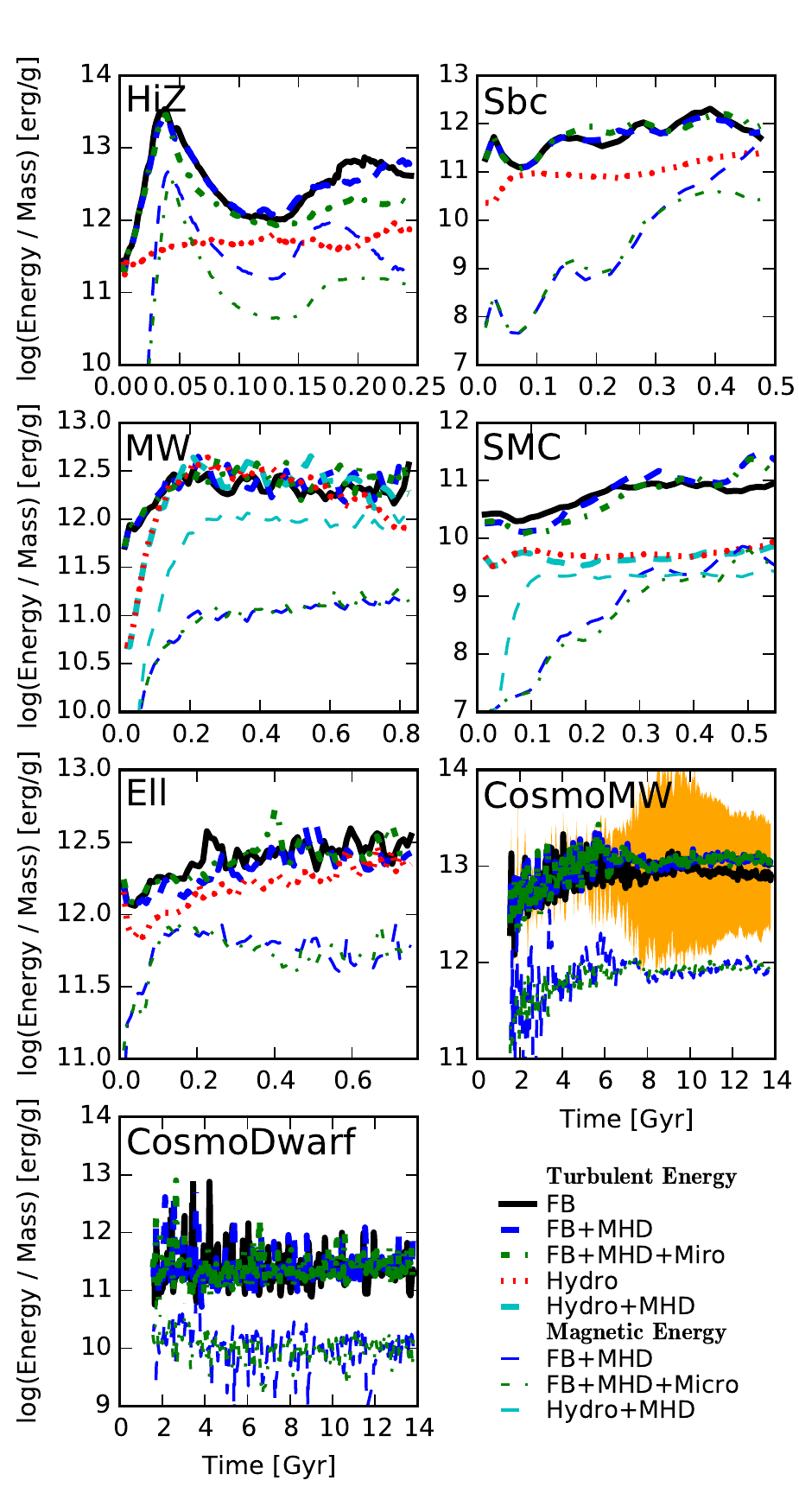}

\label{fig:turbulent}
\caption{The total turbulent kinetic energy ({\em thick lines}; defined in \sref{s:magnetic}) and magnetic energy ({\em thin lines}) per unit mass of the non-outflowing disc gas in our simulations is shown as a function of time. In the CosmoMW case, the orange shaded region shows the magnitude of stochastic effects on the turbulent energy. In all cases with feedback, the turbulent energy saturates at a steady-state value over many dynamical times, corresponding to super-sonic and  super-Alfv{\'e}nic velocity dispersions. The steady-state value is only very weakly altered by MHD and microphysical diffusion, consistent with expectations for super-Alfv{\'e}nic turbulence on large (galactic) scales. The magnetic energy grows from the seed value to $\sim 10\%$ of the turbulent kinetic energy, consistent with idealized simulations of the supersonic turbulent dynamo. Runs without feedback produce noticeably weaker turbulence  (although {\em local} bulk motions from collapsing structures can be large, they are excluded by our estimator) and equipartition magnetic energy, thus suggesting that stellar feedback participates in driving turbulence. Note that MHD+hydro runs were done only for MW and SMC. }
\end{figure}

\fref{fig:turbulent} compares the turbulent and magnetic energies in these simulations. The ``turbulent'' energy is difficult to define in practice, since we wish to exclude non-circular bulk motions and galactic winds. For our isolated galaxies, we focus on the galactic disc by taking a cylinder with radius $10\,$kpc and height $2\,$kpc centered on the disc. The cylinder is divided into annuli with thicknesses set so that the number of particles in each layer is proportional to the order of layer counted from inside out.  Within each annulus, the average rotational velocity is subtracted, and the particles outside with the highest $32\%$ $|v_z-\bar{v}_z|^2$ are excluded  (to approximate an $\pm 1 \sigma$ interval) to eliminate outflows in the z direction. Each annulus is then further divided into cells with volumes set so each of them contain roughly 15 gas particles. The dimensions of the cells are chosen so that if all gas particles were distributed uniformly within the cylinder, each cell would be a cube ($\ell_r=\ell_z=\ell_\theta$) with an identical volume. Within each cell, the average velocity in $\hat{r}$, $\hat{z}$ and $\hat{\phi}$ directions are subtracted, and particles with the highest $20\%$  $|\mathbf{v}-\bar{\mathbf{v}}|^2$ are excluded to reduce the contamination from the remaining high-velocity tail resulting from outflows in all directions. The kinetic energy that remains defines our estimated ``turbulent'' energy.  
\footnote{For the HiZ runs, a cylinder with radius $35\,$kpc and height $10\,$kpc is used instead because its star-forming regions are spread over a significantly larger volume than in the other runs.} 

For CosmoMW, which has no well-defined disc structure until $z \lesssim 0.6$, and and CosmoDwarf, which has no disc structure at all, we take all particles within 0.1 virial radius into account. The 0.1 $R_{\rm vir}$ sphere is divided into shells with thicknesses set such that the number of particles within each shell is proportional to the square of the layer number counted from inside out. The total angular momentum of each shell is calculated and used to define the z direction of the corresponding shell. Each shell is then further divided into several annuli at different $\theta$, with heights set such that the number of particles in each annulus is proportional to the corresponding $\sin{\theta}$ value. The average rotational velocity of each annulus is then subtracted. After this, each annulus is separated into cells containing roughly 15 particles. The dimensions of the cells are consistently set so that if all gas particles were distributed uniformly within the 0.1 $R_{\rm vir}$ sphere, each cell would be a cube with an identical volume.  Within each cell, the average velocities in the $\hat{r}$, $\hat{\theta}$ and $\hat{\phi}$ directions are subtracted, and the particles with the highest $20\%$  $|\mathbf{v}-\bar{\mathbf{v}}|^2$ are excluded. The turbulent kinetic energy is then calculated as the remaining kinetic energy.

To avoid biasing our comparison, we calculate the volume-integrated magnetic energy only for the gas particles kept in the turbulent energy calculation. The comparisons of the resulting turbulent energy and magnetic energy per unit mass are shown in \fref{fig:turbulent}. The turbulent kinetic energy grows almost immediately in these runs and quickly reaches a quasi-steady-state saturation level. With standard feedback, the turbulent energies per unit mass of Sbc, Mw and Ell all saturate to roughly $1\times 10^{12}-3\times 10^{12}$ erg/g, corresponding to an rms turbulent velocity of $7-13 $ km/s. HiZ on the other hand has slightly higher turbulent energy,  $3\times 10^{12}-6\times 10^{12}$ erg/g, corresponding to an rms turbulent velocity of $15-20 $ km/s. Among all the runs, the SMC and CosmoDwarf runs have the lowest turbulent energy ($1\times 10^{11}-3\times 10^{11}$ erg/g) and  rms turbulent velocity (2-4 km/s), owing to this galaxy having a significantly lower mass and thus requiring less turbulent energy to self-regulate \citep{2013MNRAS.433.1970F,2015arXiv151005650H}.  The turbulent energy of CosmoMW at low redshift is calculated to be roughly the same as the values of HiZ, which is higher than the results from the isolated MW simulations. However, the turbulent energy at low redshift may be slightly over-predicted since all the particles within 0.1 virial radius are included for consistency even though a disc is already formed. If we include only the gas particles in the disc at low redshift, the turbulent energy drops to roughly $1\times 10^{11}-3\times 10^{11}$ erg/g, similar to the results from the isolated MW simulations. The values we get are in good agreement with observations \citep{1992AJ....103.1552M,2006ApJ...643..881L} and the theoretical prediction that turbulent velocity for a marginally stable (turbulent $Q\sim 1$) disc is  $\sigma_T \sim f_{\rm gas} v_c$, where $v_c$ is the circular velocity, and $f_{\rm gas}$ is the ratio of the thin-disc gas mass to total enclosed mass in the galaxy \citep{2013MNRAS.433.1970F,2015arXiv151005650H}.

In the HiZ, Sbc, and SMC simulations, the turbulent energy is considerably (a factor of $3-10$, corresponding to a factor of $2-3$ in the rms turbulent velocity) higher in the presence of stellar feedback than when stellar feedback is absent, regardless of whether microphysical processes are included. In the Ell and MW simulations, the differences are much weaker - this is merely because these are our only two gas-poor galaxies ($f_{\hbox{gas}} \lesssim 0.1$). In this case, pure gravitational effects (accretion, spiral arms, etc.) can easily drive sufficient turbulent velocities to reach $Q\sim 1$ where we see the velocities saturate. 

\begin{figure}
\centering
\includegraphics[width=8.4cm]{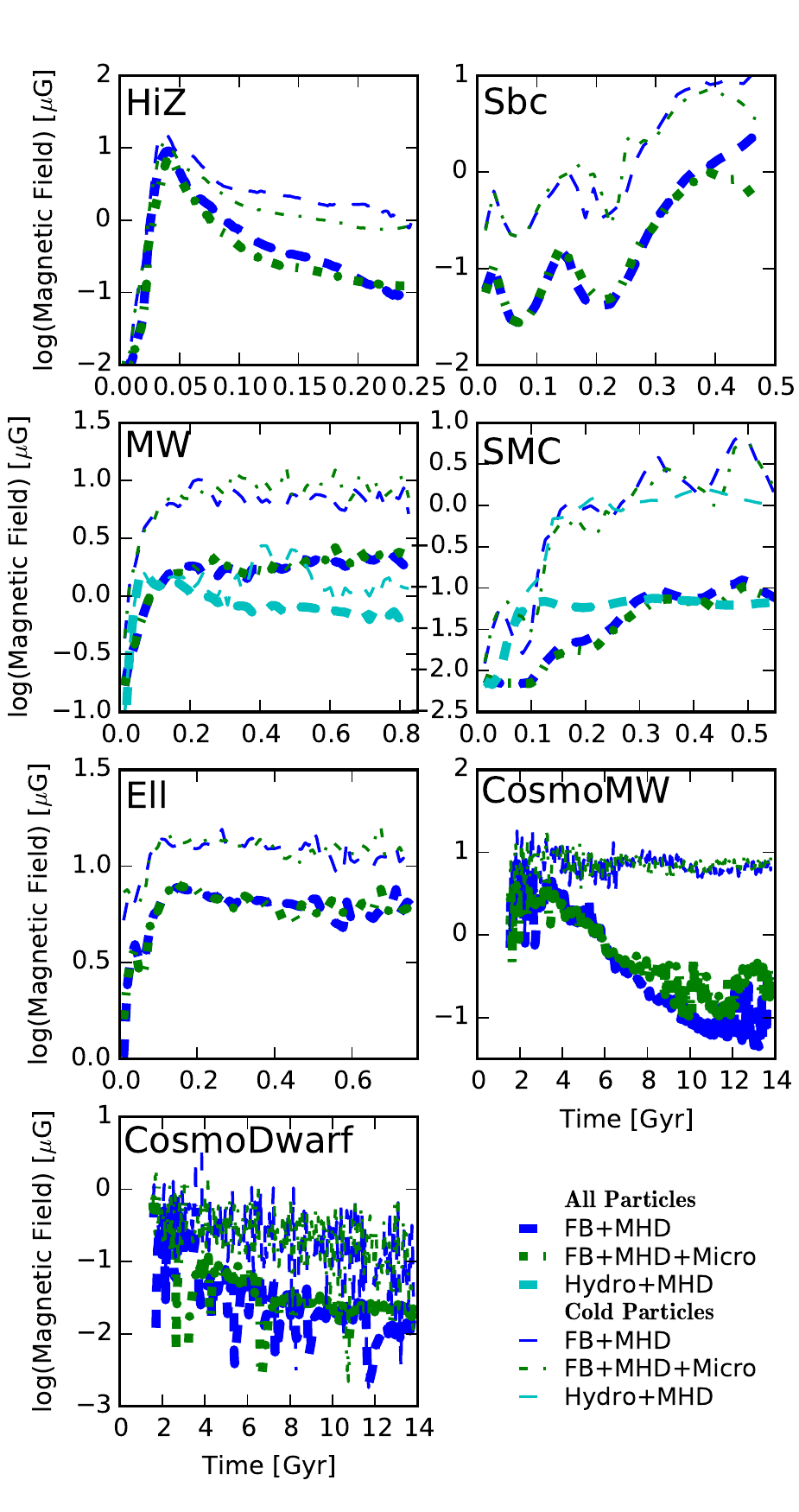}

\label{fig:magnetic_growth}
\caption{The rms magnetic field strength of all ({\em thick lines}) and the $T < 8000$ K component ({\em thin lines}) of the non-outflowing disc gas in our simulations is shown as a function of time.  Although the average value over all particles varies among different galaxies, 
the rms magnetic field strength of cold gas saturates at several to 10 $\mu G$ except CosmoDwarf, consistent with observations and other simulations. In the CosmoDwarf case, the saturation value of the cold-gas magnetic field strength is only 0.1-1 $\mu G$, owing to the frequent violent blowouts of the galaxy's ISM.}
\end{figure}

As expected, we see that the magnetic energy grows from being negligible relative to the turbulent energy (because of the small initial seed fields used) until it saturates at $3\times 10^{10}-5\times 10^{11}$ erg/g in the HiZ, Sbc, MW and Ell runs and $\sim 3\times10^{9}-10^{10}$ erg/g in the SMC and CosmoDwarf runs; these values are roughly $10\%$ of the turbulent energy similar to values measured in idealized simulations of the super-sonic turbulent dynamics \citep{2009ApJ...696...96W, 2010A&A...523A..72D,2010ApJ...716.1438K}.  However, since (by construction) the initial field value is close to the equipartition value in the Ell run, the amplification is relatively mild in this case. Conduction, viscosity, and turbulent metal diffusion have little effect on the saturated field strengths. The corresponding volume-weighted rms magnetic fields are shown in \fref{fig:magnetic_growth}, where the thick lines show the rms magnetic fields of all gas particles and the thin lines show the values of only the cold ($<8000$K) gas\footnote{To suppress noise in the mean cold gas magnetic field, we exclude gas with density less than $10^{-6}$ cm$^{-3}$ for CosmoMW and $10^{-2}$ cm$^{-3}$ for HiZ.}. Although the total rms magnetic fields vary among different galaxy types because we use the same sampling volume for galaxies with different size, the rms magnetic fields of cold particles saturates to roughly 10 $\mu G$ in all cases except CosmoDwarf, in good agreement with both observations \citep{,1996ARA&A..34..155B,2002RvMP...74..775W,2008RPPh...71d6901K,2008Natur.454..302B,2008ApJ...676...70K,2012ApJ...757...14J,2012ApJ...761L..11J} and other simulations \citep{2009ApJ...696...96W,2010A&A...523A..72D,2010ApJ...716.1438K,2011MNRAS.415.3189K,2012MNRAS.422.2152B,2013MNRAS.432..176P}.  In CosmoDwarf, the saturation value of the magnetic field in the cold gas is smaller, 0.1-1$\mu G$, perhaps because essentially the entire ISM is blown out multiple times over the course of the simulation; consequently, the time for which a given parcel of cold gas remains in the disc and has its field amplified via differential rotation and the small-scale turbulent dynamo is shorter than in e.g. the MW case. In the MW Hydro+MHD case, although the magnetic energy per mass is higher than in rest of the runs, the volume-weighted rms magnetic field strength is lower because the dense gas fraction is lower than in e.g. HiZ.

The direct comparison of magnetic and turbulent energy clearly illustrates that the turbulence is both super-Alfv{\'e}nic and super-sonic. In this limit, we expect magnetic fields to have a negligible effect on the turbulent kinetic energy and only a weak effect on the density fluctuations driven by turbulence \citep{2012arXiv1203.2117M,2008ApJ...688L..79F,2011ApJ...731...62F,2013A&A...549A..53K}. In contrast, as shown explicitly in the MW and SMC Hydro+MHD runs, the turbulent energy and magnetic energy of the runs without feedback reach approximate equipartition because both the turbulent and magnetic energy are concentrated in dense clumps and driven by gravitational collapse. In this case, magnetic fields have a stronger back-reaction on the turbulent flow, and the turbulence is therefore no longer isotropic \citep{2012arXiv1203.2117M}. This partially explains why magnetic fields have been observed to have strong effects in other studies where stellar feedback is absent or weak but not in ours. However, we also want to add the caveat that the balance of energy in runs without feedback could be affected by the ICs, as mentioned above.

\begin{figure}
\centering
\includegraphics[width=8.3cm]{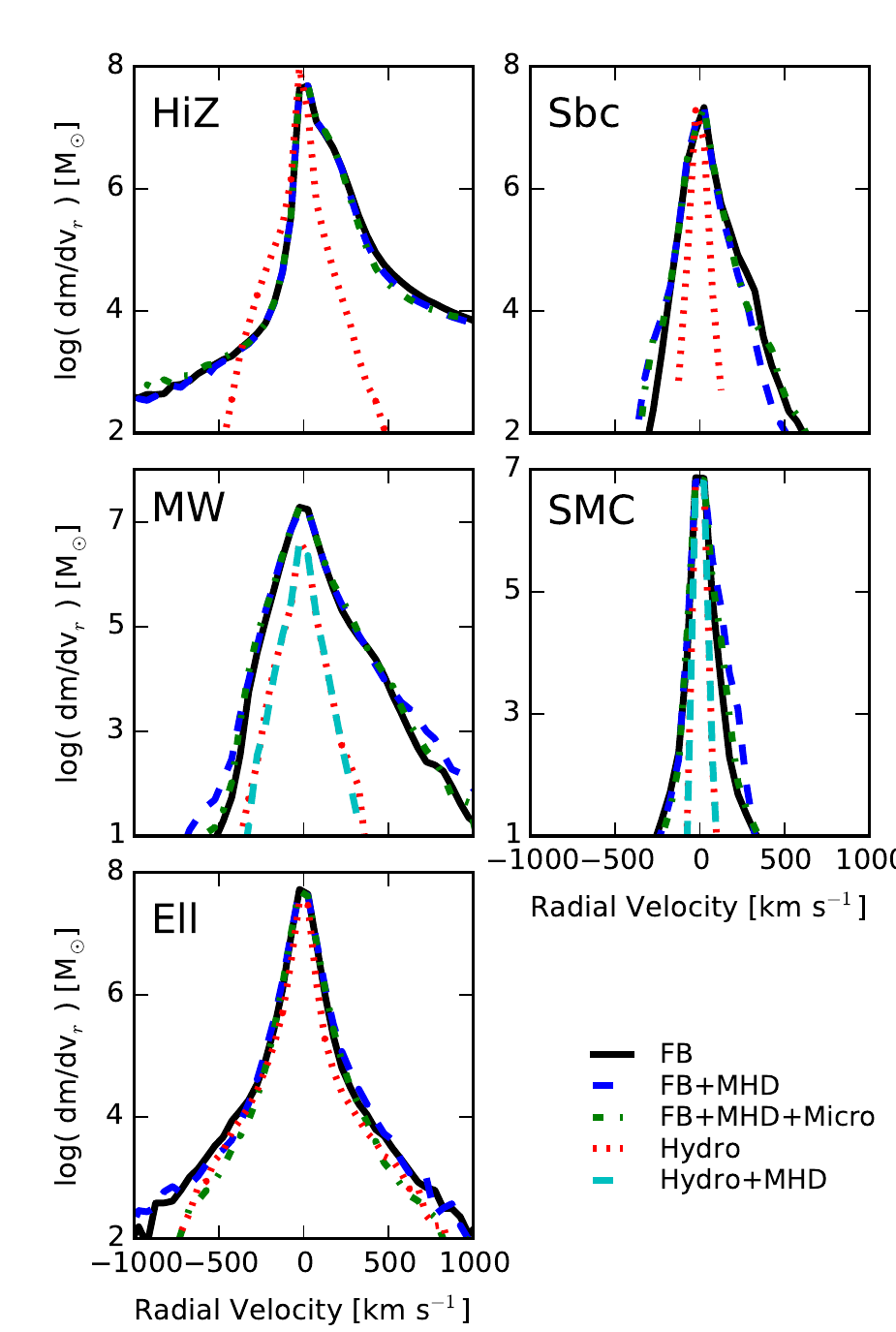}

\label{fig:wind_velocity}
\caption{Distributions of the radial velocities of the gas particles in the isolated galaxy simulations. For each, we plot the time-averaged distribution of mass as a function of the radial velocity $v_{r}$ with respect to the galaxy center of mass.  Without feedback, there are little outflows, despite the slight boost from magnetic field in the MW and SMC Hydro+MHD runs. Once stellar feedback is included, outflows are self-consistently driven (i.e. there are substantially more particles with large radial velocities) and are insensitive to the microphysics investigated here. 
}
\end{figure}

\begin{figure}
\centering
\includegraphics[width=8.3cm]{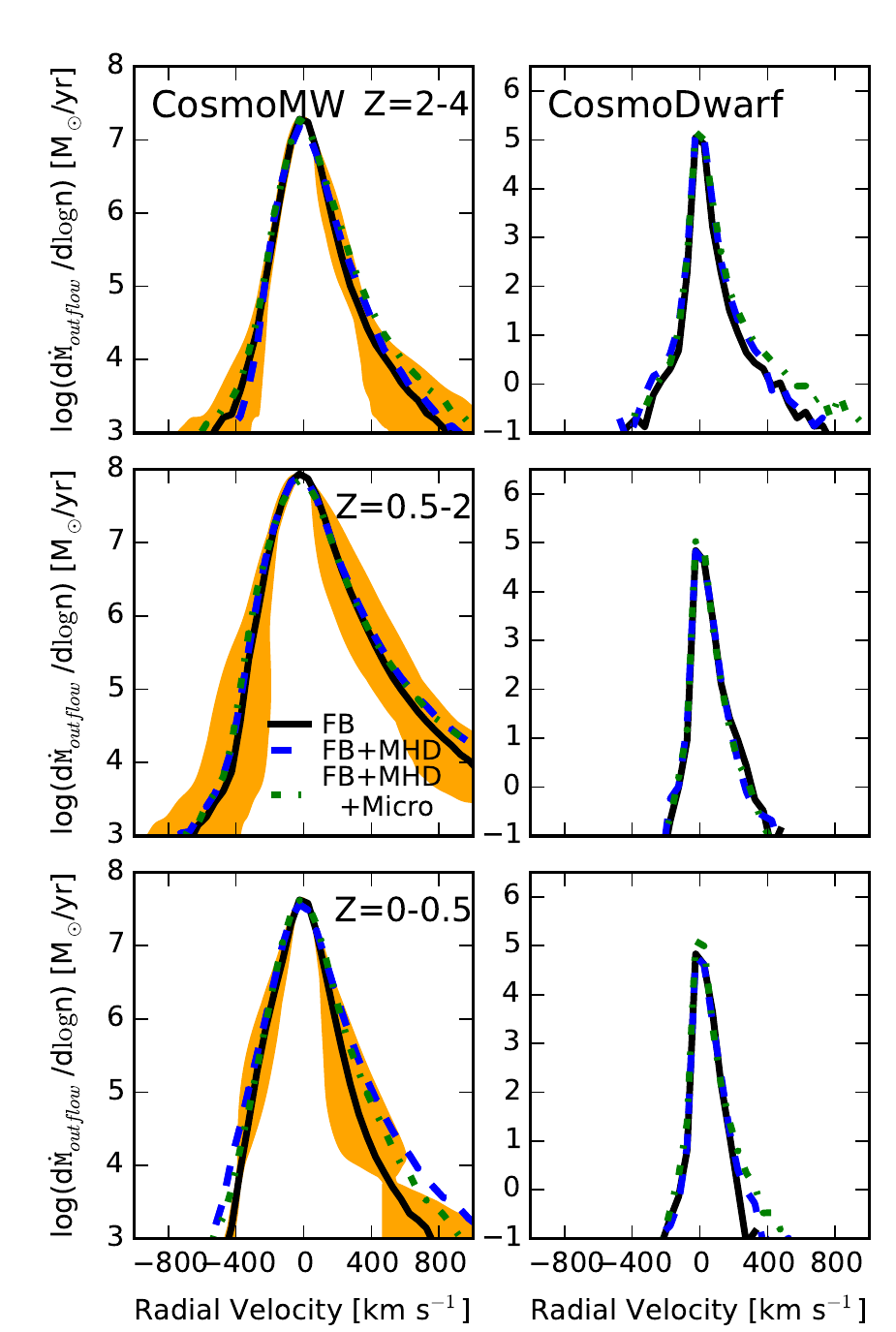}

\label{fig:wind_velocity_cos}
\caption{Distributions of the radial velocities of the gas particles in the cosmological simulations. For each, we plot the time-averaged distribution of mass as a function of the radial velocity $v_{r}$ with respect to the galaxy center of mass. The orange shaded regions in the CosmoMW panels represent the magnitude of stochastic effects. The velocity distribution function is insensitive to the microphysics investigated here. 
}
\end{figure}


\vspace{-0.5cm}
\subsection{Galactic outflows}\label{S:outflow}

In simulations that explicitly include multiple stellar feedback channels, such as those presented in this work, strong outflows are generated self-consistently. We quantify the outflow phase structure in \fref{fig:temperature}- \ref{fig:temperature_cos} and gas velocity distribution in \fref{fig:wind_velocity}. Detailed analyses of these properties using similar simulations with the same physics as our ``FB'' case are presented in \citet{2012MNRAS.421.3522H,2013MNRAS.433...78H} (for isolated galaxies)  and in \citet{2015MNRAS.454.2691M} (for cosmological simulations); here, our focus is only on how the outflow properties depend on the included microphysics.

\fref{fig:wind_velocity} and \fref{fig:wind_velocity_cos} plot the distribution of the radial velocities (defined relative to the baryonic center-of-mass of the galaxy) of gas particles within the same volume as stated in \sref{s:magnetic}\footnote{For CosmoMW, unlike in \sref{s:magnetic}, in which the same sampling volume is used for all redshifts, we switch to a cylinder with radius $10\,$kpc and height $2\,$kpc at low redshift because it better captures the wind properties at the time when a disc has already formed.} averaged over the entire simulation duration (or appropriate redshift ranges for the cosmological runs). To isolate ``outflows'' in \fref{fig:temperature} and \fref{fig:temperature_cos}, we simply take all gas particles within a thin layer at the boundary of the galaxy region defined above that have a radial velocity greater than some $v_{\rm min}$ chosen to be an appreciable fraction of the escape velocity of each galaxy ($v_{\rm min}= 200,\,100,\,100,\,30,\,100,\, 100,\, \hbox{and } 30 \,{\rm km\,s^{-1}}$ for the HiZ, Sbc, MW, SMC, Ell, CosmoMW and CosmoDwarf runs, respectively). 

Galactic outflows driven by magnetic field via Blandford-Payne-type acceleration \citep{1982MNRAS.199..883B} are visible in the absence of feedback, as shown in the Hydro+MHD runs in both the MW and SMC cases. However, it turns out that the wind mass loading owing to Blandford-Payne acceleration is orders of magnitude smaller than what is caused by feedback and therefore not evident in FB+MHD runs.  In sum, MHD and the additional physics that we study appear to have little effect on the velocity or density of the outflows compared with the effects of feedback. 


\vspace{-0.5cm}
\section{Discussion: why are the effects of the additional microphysics weak?} \label{s:discussion}
We have seen systematically that for the large-scale properties of star formation, ISM structure, and galactic outflows, magnetic fields and microphysical diffusion processes make little difference once explicit stellar feedback is included. Here we discuss why this is the case.

\vspace{-0.5cm}
\subsection{Sub-grid metal diffusion from unresolved turbulence}\label{s:metal}

The microscopic (Brownian) diffusivity of metals is negligibly small for the systems we simulate. Instead, Smagorinski-type (mixing length theory) models assume that unresolved turbulent eddies can be treated as a scale-dependent diffusion process with diffusivity $\sim v_{t}(\lambda)\,\lambda$ (where $v_{t}(\lambda)$ is the rms turbulent velocity measured on length scale $\lambda$). The ``sub-grid'' part of the model applies an explicit extra diffusion term using $\lambda\rightarrow \Delta x$ (where $\Delta x$ is the spatial resolution) to account for un-resolved eddies, with the assumption that all larger eddies are resolved and that the $v_{t}(\Delta x)$ measured around each point is indeed turbulent motion. But for a super-sonic cascade, or any system obeying the observed line width-size relation \cite[which we have demonstrated is satisfied by our simulations with stellar feedback in][]{2012MNRAS.421.3488H}, $v_{t}(\lambda)\,\lambda\sim  (G \Sigma \pi)^{1/2}  \lambda^{3/2}$, where $\Sigma$ is the surface density. Thus, the power and diffusivity is concentrated in the largest-scale eddies. In the disc, these have scales of order the disc scale height; in the halo, the relevant scales are a large fraction of the halo core radius. In either case, the largest-scale eddies are well-resolved. In fact, taking the resolved line width-size relations found in our previous work \citep{2012MNRAS.421.3488H} and integrating (assuming an infinite inertial range), we expect that most of the global diffusivity is resolved. Thus, the only effect of the sub-grid diffusivity is to smooth particle-to-particle variations in metallicity after bulk mixing of the metals (by particle motion) is resolved. In principle, this can alter the cooling rates, but the effect is weak because the resolution is high enough that we account for individual SNe explosions (so that the Poisson noise in the number of enrichment events that each particle sees is small; \citealt{2015arXiv150402097M, 2015MNRAS.447..140V}). Although subtle effects, e.g.\ the predicted dispersion in abundances within star clusters, may not be well captured by our simulations, these have little effect on the global properties that we focus on herein. 

To the extent that much larger effects are found using similar mixing models \citep{2010MNRAS.407.1581S,2009MNRAS.399..574W}, one of three effects is likely to be occurring in those works. (1) The turbulent driving scales are not resolved (so there is little or no resolved mixing). This is certainly the case in simulations with force softening $\gtrsim100\,$pc. However, in this case, it is not correct to apply the Smagorinski model in its typical form, since (as discussed above) it explicitly assumes that all shear motion in the simulation around a particle is resolved turbulent motion (from which it extrapolates the inertial range). (2) Additional motions (e.g.\ orbital motions in a disc or outflow motions in winds) are accidentally triggering the numerical ``turbulent velocity'' estimator. This effect is also likely to be more severe in lower-resolution simulations but is a serious concern when sub-grid diffusion models are applied to galaxy simulations at any resolution. (3) The coefficient of the diffusivity is too large (or the numerical gradient estimator is inaccurate), so the diffusivity assigned to un-resolved eddies can be larger than that of larger, resolved eddies. This can easily occur if $C\gtrsim1$ is used, or if the gradient estimator is noisy  (which is commonly the case in SPH).
 
We note that the mean metallicity and the vertical metallicity gradient are not affected by sub-grid turbulent metal diffusion \citep{2016arXiv160804133M}. This and the fact that the initial metallicity of some ICs is already high also help suppress the effect of turbulent metal diffusion on the properties investigated. However, when sub-grid turbulent metal diffusion is included, the metallicity PDF become much narrower because most of the gas particles eventually reach the mean metallicity value in some local annulus of the disc and therefore so do the star particles spawned from gas particles. A detailed study of the stellar metallicity distribution with and without sub-grid turbulent metal diffusion and comparisons with observational constraints will be included in \citet[][in preparation]{2016AAS...22820901E}.

In this study, we only apply the sub-grid model for unresolved turbulence to metal diffusion. In principle, unresolved turbulence can also cause diffusion of the other quantities, such as energy or momentum. However, unlike metallicity (which is not advected across particles in the MFM approach or in e.g. SPH), these quantities are readily exchanged between particles through the hydrodynamic equations and therefore have the inherent numerical diffusion from our Godunov-type solver that scales (crudely) as $\sim c_{s}\,\Delta x$ (where $\Delta x$ is the spatial resolution). As a result, the corresponding small-scale diffusion is not as significantly underestimated as metal diffusion when sub-grid turbulent diffusion is omitted. Moreover, in the sub-grid model here, the magnitude of unresolved turbulence is estimated through the sheer tensor, whereas the SGS model \citep{2006A&A...450..265S,2006A&A...450..283S} may possibly be a more rigorous approach, especially when the motion is highly shearing \cite{2016arXiv161006590C}.

\vspace{-0.5cm}
\subsection{Conduction \&\ viscosity}
We can also understand why physical conduction and viscosity have weak effects in the simulations presented in this work. 
Like all numerical methods, our Godunov-type solver has inherent numerical diffusion, with a numerical diffusivity $\sim c_{s}\,\Delta x$. 
Comparing this to Spitzer-Braginskii conduction (diffusivity $\sim (\kappa\,m_{p})/(k_{B}\,\rho)$), and using the fact that our code is Lagrangian (so $\rho\sim m_{i}/\Delta x^{3}$, where $m_{i}$ is the particle mass), we estimate that physical conductivity dominates  our numerical when $T \gtrsim 0.3\times 10^{7}\,{\rm K}\,(m_{i}/10^{4}\,M_{\sun})^{1/6}\,(n/0.01\,{\rm cm^{-3}})^{1/3}$. Moreover, as mentioned in \sref{s:metal}, large-scale turbulent eddies have diffusivity  $\sim v_{t}(\lambda)\,\lambda \sim (G \Sigma \pi)^{1/2} \lambda^{3/2}$.  We estimate the micro-physical conductivity to surpass this turbulent value at an even higher temperature $T\gtrsim 3\times 10^7 \mathrm{K} (\Sigma/(100 M_{\odot} pc^{-2}))^{3/5}(\lambda/100 pc)^{1/5}$. Thus, only in the hot, tenuous gas phase can physical conductivity be significant relative to the turbulent (and numerical) diffusivity. 
The Field length, i.e. the characteristic scale below which conduction is faster than cooling, under these conditions is $\sim [(\kappa\,m_{p}\,t_{\rm cool})/(k_{B}\,\rho)]^{1/2} \sim {\rm 10 kpc}\,(T/10^{7}\,{\rm K})^{7/4}\,(n/0.01\,{\rm cm^{-3}})^{-1}$. In hot gas, this scale is resolved. However, for typical ISM temperatures ($\sim 10^4 K$), which is the relevant regime for clump formation and star formation, the Field length is sub-pc.
In our simulations, this is typically much smaller than the resolved scales of structures (including coherent filaments and dense gas blobs). 
This indicates that while conduction and viscosity may have interesting effects on small scales, explicitly including them in galaxy simulations at present is not expected to have a large impact. 
This does not, of course, guarantee that conduction and viscosity cannot have effects that feed back to large scales in fully resolved simulations. 
More work will be needed to answer this latter question.

The only exception, where conduction and viscosity generate a small, but visible systematic effect is the CosmoDwarf case, in which the FB+MHD+Micro run has $\sim 0.2$ dex higher stellar mass and cold gas mass and more stable SFR and cold clumps. These are most likely caused by conduction. In a dwarf galaxy of this mass, an overlap of SNe can easily wipe out all the cold gas in the galaxy, which happens several times throughout the simulations. However, conduction could dissipate part of the SNe energy, making it more difficult for them to completely destroy the cold clumps in the ISM. From the star formation histories, we infer that a typical event capable of ``wiping out'' all the cold gas requires an overlap of  $\sim100$ SNe. To show that conduction can actually be effective in this limit, we can compare the time scale of energy dissipation by conduction,   $\tau_c\equiv E/ \dot{E}_{\text{conduction}}  $, to the free expansion time of SNe in the energy conserving phase,   $\tau_{\text{exp}}\equiv R/v $, where $E=N_{\text{SNe}}10^{51}$ erg  is the energy sum of the $N_{\text{SNe}}$ overlapping SNe, $R$ is the radius of the expanding hot bubble and $v$ is the speed of expansion. $\tau_c/\tau_{\text{exp}}$
 turns out to be $\sim 6\times 10^{-6} (R/10pc)^7 (N_{\text{SNe}}/100)^{-2} (n/cm^{-3})^3<1$, which means conduction cannot be neglected, (see also \cite{2014MNRAS.442.3013K,2015MNRAS.453.3499K}). As the expansion continues, the radius grows, $\tau_c/\tau_{\text{exp}}$ increases and  the importance of conduction gradually decays. However, before conduction ceases to be important ($\tau_c/\tau_{\text{exp}}=1$), the hot bubble can entrain a total mass of $\sim 10^4 (N_{\text{SNe}}/100)^{6/7} (n/cm^{-3})^{-2/7} M_\odot$, which is well resolved in our CosmoDwarf simulations. The effects of conduction are therefore expected to be visible in our CosmoDwarf simulations, where the resolution is high enough and the size small enough for the cold gas to be mixed in overlapping SNe remnants. 

\vspace{-0.5cm}
\subsection{Magnetic fields}
\label{s:discussion.bfield}
We can also understand in simple terms why magnetic fields have little effect on the galaxy SFRs and, consequently, their position in the Schmidt-Kennicutt diagram and other SF scaling laws. In a series of previous studies \citep{2011MNRAS.417..950H,2012MNRAS.421.3488H,2013MNRAS.430.1901H,2013MNRAS.433...69H,2013MNRAS.432.2647H,2013MNRAS.433.1970F,2014MNRAS.445..581H}, we have repeatedly shown that galaxy SFRs are set by a balance between stellar feedback injecting momentum (``resisting'' collapse and the``stirring'' of super-sonic turbulence) and dissipation of that motion via gravitational collapse. Other groups have reached consistent conclusions in calculations that include self-gravity, resolve fragmentation and turbulence, and explicitly model stellar feedback \citep{2008ApJ...684..978S, 2013ApJ...770...25A, 2013ApJ...776....1K,2014ApJ...786...64K}. 
In such simulations, galactic SFRs are independent of the sub-grid model for how dense gas turns into stars \citep{2011MNRAS.417..950H,2013MNRAS.433...69H}, the shape and orders-of-magnitude variations in the cooling function and modeled dense gas chemistry \citep{2012MNRAS.421.3488H}, and the temperature and detailed kinematics of the star-forming gas \citep{2013MNRAS.432.2647H}. 
Even equipartition magnetic fields would change the equilibrium SFR in this scenario by only tens of percent, and the effect should be even weaker in the super-Alfv{\'e}nic case when feedback in present.

Comparing our results with those of galaxy simulations with weaker/non-explicit stellar feedback presented in the previous literature \citep{2009ApJ...696...96W,2013MNRAS.432..176P,2012MNRAS.422.2152B}, the SFR difference caused by inclusion of magnetic fields is small even in our runs without feedback. In \cite{2013MNRAS.432..176P,2012MNRAS.422.2152B}, the Springel and Hernquist feedback model \citep{2003MNRAS.339..289S} is adopted, resulting in a smooth, pressure-supported ISM with a stiff ``effective equation of state''. In this case, the extra pressure support from magnetic fields could be more effective. In \cite{2009ApJ...696...96W}, instead of explicit star formation, the star formation rate is calculated from the amount of dense gas assuming a specific star formation efficiency. In this case, gas could possibly stay in high-density clumps for a longer time, thus allowing more time for magnetic pressure to build up and become effective.

Similar arguments apply to galactic outflows: the mass outflow rate is set by the momentum injected by feedback, which is divided into stirring turbulence in the disc and ejecting low-density material \citep{2011ApJ...735...66M,2014arXiv1411.1769T,2015arXiv151005650H,2016MNRAS.tmp..561M}, and ultimately limited by the depth of the potential \citep{2015MNRAS.454.2691M}. We thus do not expect the mass outflow rate to vary by large factors when magnetic fields are present.

Perhaps the most surprising result of this study is how weak the effects of magnetic fields are on the ISM phase structure. Using the same code in idealized tests, we have shown that sufficiently strong magnetic fields do produce qualitatively different behavior in fluid mixing instabilities, cloud entrainment or ``crushing'' by ambient winds, and anisotropic super-sonic turbulence \citep{2016MNRAS.455...51H}, in agreement with a vast literature. Moreover, the rms fields we predict in both ambient gas and dense clouds are in reasonable agreement with those observed, as discussed in \sref{s:magnetic}. However, it appears that two main effects lead to relatively weak effects of magnetic fields on the large-scale phase structures we consider. (1) The turbulence in most of the disc is super-Alfv{\'e}nic, at least on large scales that contain most of the power (of order the disc scale height, which also corresponds to the size of the largest GMCs, which contain most of the dense gas mass and star formation). Thus, the turbulent velocity dispersions, isotropy, and density fluctuations generated are not strongly altered \citep{2012arXiv1203.2117M,2008ApJ...688L..79F,2011ApJ...731...62F,2013A&A...549A..53K}. (2) GMCs are not steady-state, pressure-confined, equilibrium objects in the simulations. We have previously shown \citep{2012MNRAS.421.3488H} that in similar simulations, GMCs form rapidly (in a single dynamical time) from gravitational instability (and are self-gravitating) and live just a few dynamical times, forming stars via turbulent fragmentation until feedback disrupts the cloud.

\cite{2015MNRAS.449....2M} recently showed that magnetized gas clouds can survive much longer than unmagnetized ones when accelerated by supersonic hot wind.  Moreover, \cite{2016arXiv160805416A,2016ApJ...822...31B} showed that thermal conduction can also help cold clouds survive by suppressing the Kelvin-Helmholtz instability at the interface.
Although such effects are confirmed to exist in high-resolution wind-tunnel-type simulations run with the code used here (Hopkins et al. in preparation), the simulations presented in the current study suggest that magnetic fields play a less important role in the formation and survival of dense cold clouds in the ISM of galaxies, such as molecular clouds.
One important difference is that massive molecular clouds are typically self-gravitating rather than in pressure equilibrium with the surrounding ISM. 
Moreover, as explained above, in our simulations, massive molecular clouds are disrupted by stellar feedback rather than by hydrodynamic instabilities. Consequently, the aforementioned idealised tests may not represent the physical conditions relevant for simulated (or real) GMCs. As for the cold clouds in outflows, although our simulations also suggest weak effects from magnetic fields, we caution that only 2 galaxies in our study contain ``hot halos'' (Ell and CosmoMW). In our other simulations, such cloud ``shredding'' or mixing effect will be much weaker since there is little or no hot gas halo for the cold outflow gas to mix into. In the Ell and CosmoMW cases where there are hot halos,  the achievable resolutions are inevitably lower, meaning the phase structure in the outflows may not fully resolved. Besides, most of the outflow mass which reaches large radii ($\gtrsim R_{vir}$)  is hot gas in the first place - cold outflows tend to be recycled in small-scale fountains  (see \citealt{2015MNRAS.454.2691M}).  

Despite the weak role that magnetic fields play, we see a hint of a small ($<0.1$ dex) increase of stellar mass in those runs with magnetic fields. Interestingly, magnetic fields may, if anything, enhance star formation on large scales. The fact that the difference  is more obvious in the smallest galaxy suggests that this may  result from the magnetic field helping stabilizing colds clumps in the ISM, especially in small galaxies with a less stable cold phase.

\vspace{-0.5cm}
\section{Conclusions} \label{S:conclusions}
We use simulations with parsec-scale resolution, explicit treatments of stellar feedback identical to those used in the FIRE project, magnetic fields, anisotropic Spitzer-Braginskii conduction and viscosity, and sub-grid turbulent metal diffusion to study how these affect galaxy-scale star formation, the phase structure of the ISM, and the generation of galactic outflows. We consider both isolated (non-cosmological) simulations of a range of galaxy types and fully cosmological zoom-in simulations of a Milky Way-mass halo and a dwarf halo. 

In all cases, we find the following:

\begin{itemize}

\item{Stellar feedback plays the dominant role in regulating the SFR. 
We find that  magnetic fields and additional microphysical diffusion processes change the SFR (and therefore the KS law) by small amount comparing to the effect from stellar feedback in the investigated ma. This is consistent with the models advocated in the aforementioned papers  (see the references in \sref{s:discussion.bfield}), in which the SFR and star formation scaling relations are set by self-regulation via feedback, which drives super-sonic turbulence and balances the disc against gravity.}

\item{The ISM phase structure and galactic winds is also primarily established by stellar feedback. 
Stellar feedback also serves as an extra source of turbulent energy, boosting the rms turbulent velocity by a factor of 2-3. Perhaps surprisingly, however, neither MHD nor the additional diffusion microphysics appear to produce larger than $\sim 10\%$-level systematic effects on these quantities. In fact, in some earlier experiments where we artificially increased the viscosity coefficient $\eta$ by a factor of 100, there were still weak systematic effects. It appears that because the turbulence is super-Alfv{\'e}nic on the scales most important for fragmentation, ISM phase structure and outflow generation (of order the disc scale height), these effects are sub-dominant. A more detailed discussion of why such small effects are seen is provided in \sref{s:discussion}.}

\item{The magnetic field energies saturate at  $\sim 10\%$ of the turbulent kinetic energies on of order the galactic scale height (\fref{fig:turbulent}). The ratio is smaller still if we include the kinetic energy of small-scale galactic fountains in the ``turbulence'' budget. This is consistent with both observations \citep{,1996ARA&A..34..155B,2002RvMP...74..775W,2008RPPh...71d6901K,2008Natur.454..302B,2008ApJ...676...70K,2012ApJ...757...14J,2012ApJ...761L..11J}  and other simulations \citep{2012MNRAS.422.2152B,2013MNRAS.432..176P,2009ApJ...696...96W,2010A&A...523A..72D,2010ApJ...716.1438K,2011MNRAS.415.3189K}. This result partially explains why the magnetic field's effects are sub-dominant on the large scales of order the disc scale height (the scales containing most of the turbulent energy).}

\item{A systemic increase of stellar mass and cold gas is observed in CosmoDwarf run with all fluid microphysics included. This may result from conduction dissipating part of the SNe energy making it more difficult to wipe out  cold clumps. Our cosmoMW run shows a similar enhancement in late-time cooling from the CGM with all microphysics present. A more detailed discussion of this is provided in \sref{s:discussion}.}
\end{itemize}

It appears that, at least on galactic scales, in the presence of explicit models for multi-mechanism stellar feedback as well as self-gravity, magnetic fields and additional diffusion microphysics (such as conduction, viscosity, and sub-grid turbulent  metal diffusion) are subdominant in the star formation and galaxy formation process at currently achievable resolutions. This general result appears to contradict some earlier claims in the literature. However, to our knowledge, these prior studies have not focused on the combination of large galactic scales (yet with high enough resolution to resolve vertical disc scale heights and the phase structure in discs) and fully explicit models for stellar feedback. For example, it is relatively ``easy'' for magnetic fields to have a large fractional effect in simulations with either no or weak stellar feedback or stellar feedback modeled only in a ``sub-grid'' fashion (so it e.g.\ does not locally alter the gas dynamics but only ejects gas in outflows or adds an effective pressure term). However, the claimed effects in these cases are typically order-unity \citep{2005ApJ...629..849P,2007ApJ...663..183P,2009ApJ...696...96W,2012MNRAS.422.2152B,2013MNRAS.432..176P} and thus still orders of magnitude less than the factor $\sim 100-1000$ changes in the properties we study here that occur when the full model for stellar feedback is introduced. 

Altogether, our results support the emerging picture wherein galaxy-scale ($\gtrsim 10-100\,$pc) star formation, ISM structure, and outflows are determined primarily by a competition among super-sonic (and super-Alfv{\'e}nic) turbulence, stellar feedback, and self-gravity. The microphysics we study here may certainly be important on smaller scales (e.g. for regulating the structure of turbulent cores as they collapse to form stars) or in the more diffuse CGM and IGM (e.g. the outskirts of galaxy clusters). However, they do not, to leading order, significantly alter the dynamics on the scales we study here. We also caution that certain unresolved processes (e.g. conduction altering mixing and cooling in single SNe blastwaves or cool cloud ``shredding'' in the circumgalactic medium) may have large non-linear effects on the efficiency of feedback or cooling, and these cannot be captured in our simulations. We see tentative evidence of this in our fully cosmological MW-mass simulation, which shows enhanced late-time cooling and a larger gas disc with conduction, viscosity and sub-grid metal diffusion active.
 
Although the magnetic field has little effect on the properties analyzed in our current study, it might for instance provide important pressure support in the violent tidal compression that occurs in galaxy mergers, which could possibly affect the properties of the star clusters formed in merger-induced starbursts. Besides stellar feedback and fluid microphysics, AGN feedback may be an important determinant of galaxies' physical properties, especially for massive galaxies. Moreover, cosmic rays may significantly affecting galaxy evolution, and properly treating cosmic ray transport requires an accurate determination of the magnetic field.  Detailed investigations of these processes and their interaction with fluid microphysics in the context of simulations with explicit stellar feedback will be presented in future work.


\vspace{-0.7cm}
\acknowledgments
We thank Ai-Lei Sun, Shu-heng Shao, Eliot Quataert and Cameron Hummels for useful discussions.
Support for PFH was provided by an Alfred P. Sloan Research Fellowship, NASA ATP Grant NNX14AH35G, and NSF Collaborative Research Grant \#1411920 and CAREER grant \#1455342. CCH is grateful to the Gordon and Betty Moore Foundation for financial support. The Flatiron Institute is supported by the Simons Foundation. CAFG was supported by NSF through grants AST-1412836 and AST-1517491, by NASA through grant NNX15AB22G, and by STScI through grants HST-AR- 14293.001-A and HST-GO-14268.022-A.
DK was supported in part by NSF grant AST-1412153. Numerical calculations were run on the Caltech compute cluster ``Zwicky'' (NSF MRI award \#PHY-0960291) and allocation TG-AST130039 granted by the Extreme Science and Engineering Discovery Environment (XSEDE) supported by the NSF. 
\\

\footnotesize{
\bibliographystyle{mn2e}
\bibliography{mybibs}
}
\appendix
\normalsize
\section{Convergence tests}\label{A:resolution}
We performed convergence tests using our isolated SMC model, varying the particle mass by 2 orders of magnitude (see \tref{tab:ic2}). 
The HR run in \tref{tab:ic2} matches the resolution of the standard SMC runs in the main text. 

\begin{figure}
\centering
\includegraphics[width=7cm]{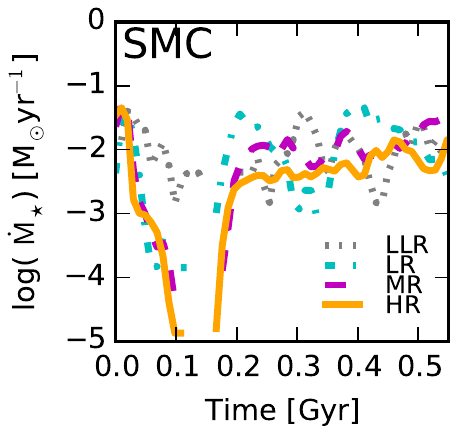}
\label{fig:sfr_div}
\caption{Convergence of the star formation rate of the SMC model. The star formation rate converges quickly. Even at a resolution 2 orders of magnitude lower than the standard resolution, the SFR has a similar quasi-equilibrium value, $\sim 0.01 M_\odot/yr$.}
\end{figure}

\begin{figure*}
\begin{flushleft}
\centering
\includegraphics[width=18cm]{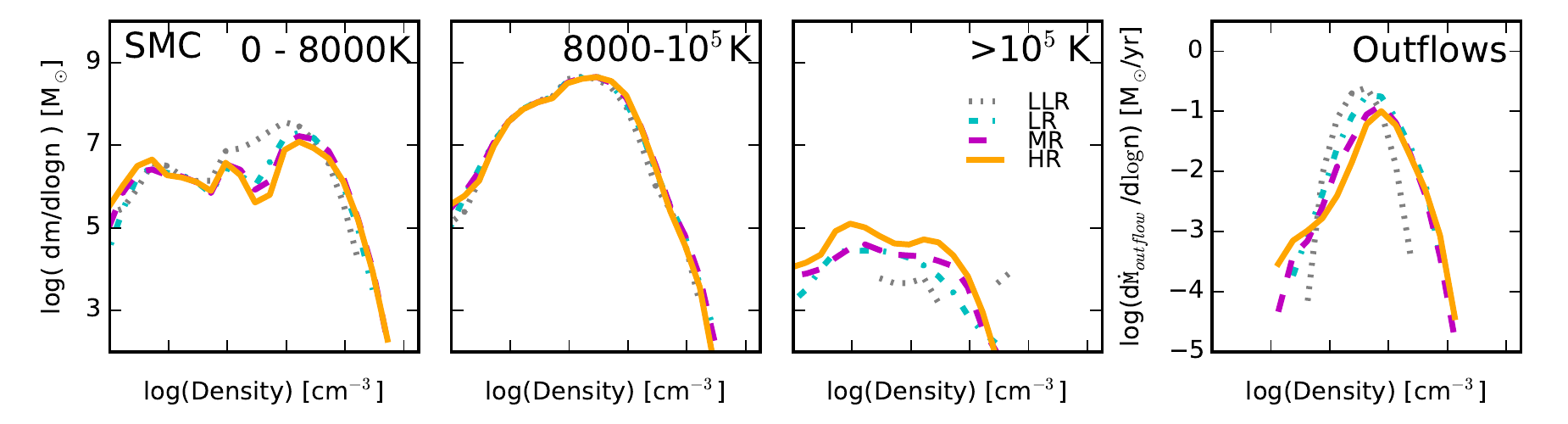}
\end{flushleft}
\label{fig:temperature_div}
\caption{Convergence of the density distribution of gas in different phases in the SMC model. The cold neutral and warm ionized gas have very similar density distributions at all of the resolutions inspected. The properties of the hot gas and outflows, on the other hand, appear to require resolution elements of $\sim 1000$ M$_{\odot}$ per gas particle, which roughly separates whether SNe are individually resolved. Nevertheless, the density distributions of the hot gas and outflows in the  lower-resolution runs do not differ drastically from the converged values.}
\end{figure*}

\begin{figure}
\centering
\includegraphics[width=7cm]{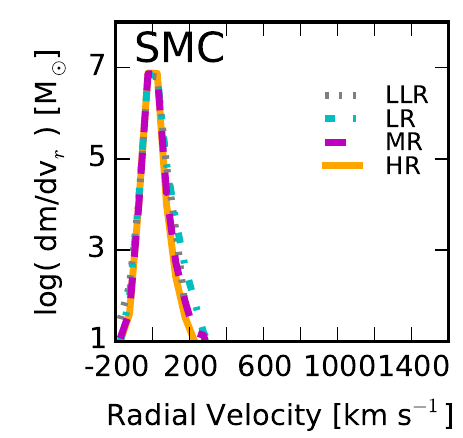}
\label{fig:wind_div}
\caption{Convergence of the radial velocity distribution of the gas particles in the SMC model. For all of the tested resolutions, the gas particles have almost identical radial velocity distributions.}
\end{figure}

\begin{figure}
\centering
\includegraphics[width=8cm]{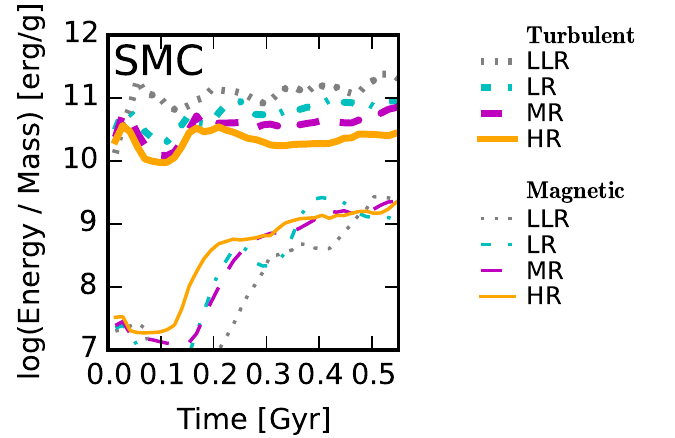}
\label{fig:turb_div}
\caption{The total turbulent kinetic energy ({\em thick lines}; defined in \sref{s:magnetic}) and magnetic energy ({\em thin lines}) per unit mass of the non-outflowing disc gas in the SMC model. Both the magnetic and turbulent energies appear converged once the gas mass resolution is $\sim 1000$ M$_{\odot}$.}
\end{figure}

The resulting SFRs are shown in \fref{fig:sfr_div}, the phase structure and radial velocity distribution are shown in \fref{fig:temperature_div} and \fref{fig:wind_div}, respectively, and the turbulent and magnetic energies are shown in   \fref{fig:turb_div}. The star formation rate in\fref{fig:sfr_div} converges most rapidly with resolution. Among the inspected resolutions, there is little difference. As for the phase structure \fref{fig:temperature_div}, the cold neutral and warm ionized gas have very similar density distributions at all the resolutions tested. The hot gas and outflow density distributions converge more slowly but appear to be converged when at the MR resolution (i.e.\ resolution elements of a few thousand solar masses), as does the radial velocity distribution of gas particles \fref{fig:wind_div}.  Above this resolution, individual SN remnants begin to have their Sedov-Taylor phases resolved, and therefore generation of hot gas and outflows can be captured more robustly. The turbulent and magnetic energies similarly appear converged at the MR resolution. As the resolution increases, minor increases in the magnetic energy and minor decreases in the turbulent energy are found. This is because the small-scale shear field, which can dissipate turbulent kinetic energy and enhance the magnetic energy through field-line stretching, is suppressed at low resolution \citep{1995ApJ...453..332J}. 

The convergence tests imply that our simulations of the more-massive galaxies, such as HiZ, Ell and CosmoMW, might not have sufficient resolution for all of their properties to be fully converged, especially their hot gas and outflow properties. 
Moreover, it is worth noting that, although we do not expect this to be the case, we cannot exclude the possibility of false convergence. The best resolution that we can achieve for galaxy simulations is inevitably many orders of magnitude coarser than the natural viscosity scale (the Kolmogorov length scale). 
Thus, it is possible that some important effects of fluid microphysics will appear only at much higher resolutions than these that will be achievable for galaxy simulations in the foreseeable future.

\begin{table*}
\begin{center}
 \caption{Galaxy models used in our convergence tests}
 \label{tab:ic2}
 \begin{tabular}{@{\extracolsep{\fill}}ccccccc}

 \hline
\hline
Initial Condition &Physics &Resolution     & $m_g$  &$m_h$ &$m_d$ &$m_b$  \\             
\hline 
SMC    &FB+MHD+Micro   & LLR                     & 3.6e4    &2.9e5    & 6.2e4  &4.8e4\\
SMC    &FB+MHD+Micro   & LR                      & 3.6e3    &2.9e4    & 6.2e3  &4.8e3\\
SMC    &FB+MHD+Micro   & MR                   & 1.1e3    &8.6e3    & 1.9e3  &1.4e3\\
SMC    &FB+MHD+Micro   & HR                    & 3.6e2    &2.9e3    & 6.2e2  &4.8e2\\
\hline
\hline
\end{tabular}
\end{center}
\begin{flushleft}
(1) Initial Condition: Galaxy model used. These all adopt our SMC IC.
(2) Physics: These all consider FB+MHD+Micro, the most demanding case.
(3) Resolution name.  LLR: The lowest resolution. LR: Low resolution. MR: Medium resolution. HR: High resolution. 
(4) $m_g$: Gas particle mass.
(5) $m_h$: Halo particle mass.
(6) $m_d$: Stellar disc particle mass.
(7) $m_b$: Bulge particle mass.
\end{flushleft}
\end{table*}

\label{lastpage}

\end{document}